\documentclass[fleqn,preprint,12pt]{elsarticle}
\usepackage{multirow}
\usepackage{amsmath}
\usepackage{amssymb}
\usepackage{amstext}
\usepackage{epsfig}
\usepackage{graphicx}
\usepackage{bm}
\usepackage{eufrak}
\usepackage{subfigure}
\usepackage{appendix}
\usepackage{xcolor}
\usepackage{algorithm}
\usepackage{algorithmic}

\def\yhat{\hat{y}}

\newcommand{\appsection}[1]{\let\oldthesection\thesection
  \renewcommand{\thesection}{Appendix \oldthesection}
  \section{#1}\let\thesection\oldthesection}

\journal{Journal of Computational Physics}

\begin{document}

\begin{frontmatter}

\title{Deep learning algorithm for data-driven simulation of noisy dynamical system }

\author{Kyongmin Yeo${}^*$}
\cortext[cor]{Corresponding author}
\ead{kyeo@us.ibm.com}

\author{Igor Melnyk }

\address{IBM T.J. Watson Research Center, Yorktown Heights, NY, USA}

\begin{abstract}
{
We present a deep learning model, DE-LSTM, for the simulation of a stochastic process with an underlying nonlinear dynamics. The deep learning model aims to approximate the probability density function of a stochastic process via numerical discretization and the underlying nonlinear dynamics is modeled by the Long Short-Term Memory (LSTM) network. It is shown that, when the numerical discretization is used, the function estimation problem can be solved by a multi-label classification problem. A penalized maximum log likelihood method is proposed to impose a smoothness condition in the prediction of the probability distribution. We show that the time evolution of the probability distribution can be computed by a high-dimensional integration of the transition probability of the LSTM internal states. A Monte Carlo algorithm to approximate the high-dimensional integration is outlined. The behavior of DE-LSTM is thoroughly investigated by using the Ornstein-Uhlenbeck process and noisy observations of nonlinear dynamical systems; Mackey-Glass time series and forced Van der Pol oscillator. It is shown that DE-LSTM makes a good prediction of the probability distribution without assuming any distributional properties of the stochastic process. For a multiple-step forecast of the Mackey-Glass time series, the prediction uncertainty, denoted by the 95\% confidence interval, first grows, then dynamically adjusts following the evolution of the system, while in the simulation of the forced Van der Pol oscillator, the prediction uncertainty does not grow in time even for a 3,000-step forecast.

}
\end{abstract}


\end{frontmatter}

\section{Introduction}
{
Data-driven reconstruction of a dynamical system has been of great interest due to its direct relevance to numerous applications across disciplines, including physics, engineering, and biology \cite{Brunton16,Jaeger04,Wang16}. In many real-world applications, we have only partial observations of a complex spatio-temporal process through a sensor network. As a results, the time series from a sensor network exhibits very complex behaviors, such as time-delay dynamics due to the finite information propagation time \cite{Ikeda80,Ma17,Mackey77,Yanchuk17}. Moreover, when measurements are made by a sensor network, the observations are corrupted by sensor noise, which makes the resulting time series a stochastic process. }

Modeling of such noisy dynamical systems has been studied extensively by using autoregressive stochastic process or state-space model \cite{Box08,Durbin12}. In order to make an inference tractable, most of the conventional time series analysis models, e.g., autoregressive moving average models or Kalman filter, make strong assumptions on the distributional property of the noise process, such as an additive Gaussian white noise, and linearize the dynamical system \cite{harvey90}. When the governing equations of the underlying dynamics are known, the extended and, later, unscented Kalman filters are proposed for nonlinear state estimations \cite{Wan00}. In geophysical data assimilation, the ensemble Kalman filter has become one of the standard approaches, because of its strength in providing a stable estimation of a high dimensional system \cite{Evensen03}. 
{
For the nonlinear filtering with known transition functions, particle filters, or sequential Monte Carlo methods, provide a very powerful tool for the modeling of non-Gaussian distributions \cite{Merwe01,Arula02}.
} 
{
While most of these nonlinear filtering models require at least a partial knowledge of the dynamical system \cite{Hamilton15}, in many problems, we do not have knowledge about the underlying physical processes, or the system is too complex to develop a model from the first principles \cite{Hamilton16}. 
}

It is challenging to reconstruct a nonlinear dynamical system without prior knowledge. There has been a significant process in such ``model-free'' approach for identification and prediction of nonlinear systems. The fundamental building block of many of the model-free approach comes from Takens's theorem \cite{Takens81}, or so called ``delay-coordinate embedding''. In a nutshell, the delay-time embedding constructs a $n$-dimensional phase space for a time-lagged data, e.g., $\bm{X}(t) = (\bm{x}(t),\bm{x}(t-\tau_1),\cdots,\bm{x}(t-\tau_{n-1}))$, where $\tau_i$ is a delay-time, and relies on a nearest neighborhood detection method. For a comprehensive review, see \cite{Wang16}. Recently, a convergent cross-mapping method is proposed to infer causality from nonlinear data \cite{Ma17,Sugihara12,Tsonis15}. In \cite{Hamilton16}, a ``Kalman-Takens'' filter is proposed, in which the delay-coordinate embedding is used as the nonlinear time-marching operator for the state vector. 
{
Instead of modeling the nonlinear transition function, Li \emph{et al.} \cite{Li14} proposed to use a reproducing kernel Hilbert space approach for the nonlinear estimation of the covariance structure in Kalman filter.
}

Recently, an artificial neural network equipped with many layers of hidden units has attracted great attention because of its strong capability in discovering complex structures in data \cite{LeCun15}. See \cite{Schmidhuber15} for a historical survey. The so-called deep learning provides a black-box model for a nonlinear function estimation and has been shown to outperform conventional statistical methods for data mining problems, e.g., speech recognition, image classification/identification. For sequence modeling, recurrent neural network (RNN) has been widely used \cite{GoodfellowBengio16}. To overcome the difficulties in learning a long-time dependency structure, a Long Short-Term Memory network (LSTM) is proposed \cite{Hochreiter97,Gers00}. LSTM uses multiple gating functions to conserve the information stored in its internal state for a longer period of time. LSTM has become one of the most widely used RNN. Jaeger \& Haas \cite{Jaeger04} proposed a variation of RNN, called echo state network (ESN) or reservoir computing. In ESN, a large number of dynamical systems, or reservoir, is randomly generated and the prediction is made by a linear combination of these dynamical systems. In the model training, only the parameters of the last layer of the network, i.e., the linear combination of the reservoir, is adjusted, which makes it much easier to train with a smaller data set compared to other RNNs. ESN has been studied and applied to many dynamical systems \cite{Inubushi17,LUKOSEVICIUS09,Pathak18}. Instead of relying on the recurrent structure of RNNs, there are approaches to explicit incorporate the time marching structure of the dynamical system, or partial differential equations \cite{Raissi18,Tompson17,Trischler16}. 

Considering its strength in learning a nonlinear manifold of the data and the de-noising capability\cite{Bengio13}, deep learning has a potential to provide a new tool for data-driven reconstruction of noisy dynamical system. While there is a large volume of literature on the application of artificial neural networks for the modeling of nonlinear dynamical systems, most of the studies consider noiseless data and / or regression problem, i.e., making a deterministic prediction given an input data. 
In the computer science literature, a few methods are proposed to extend the deterministic RNN for a prediction of probability distribution of sequential data. 
{
One of the conventional methods of making a probabilistic model is to assume the probability distribution of the data and to build an RNN, which outputs the parameters of the probability distribution, \emph{e.g.}, the mean and variance of a Gaussian distribution \cite{Bishop06, Graves14}. Recently, variational Bayes methods have become popular to naturally consider stochastic nature of time series. Fortunato \emph{et al.} \cite{FortunatoBV17} proposed a Bayesian RNN, where the parameters of the RNN are assumed to be Gaussian random variables. Bayer \& Osendorfer \cite{Bayer2014} developed a stochastic RNN by augmenting the internal states of the RNN by independent random variables. Chung \emph{et al.} \cite{chung15} proposed a variational RNN, which exploits the variational auto-encoder \cite{Kingma14} to encode the observed variability of time series. It should be noted that most of the probabilistic deep learning models also assume that the predictive posterior distribution is Gaussian. Goyal \emph{et al.} \cite{Goyal16} proposed an RNN trained with the generative adversarial network (GAN), which does not rely on the Gaussian assumption. However, the behaviors of the GAN-based methods for the modeling of nonlinear dynamical systems are not well understood.}

In this study, we present an RNN-based model for the data-driven inference and simulation of noisy nonlinear dynamical systems. 
{
While most of the previous deep learning models assume the Gaussian or a mixture of Gaussian distributions, the proposed RNN model aims to directly predict the probability density function without any assumption, except for smoothness, e.g., $C^0$ continuity.} We show that the function estimation problem can be solved by using a cross-entropy minimization via a numerical discretization. 
The temporal evolution of the probability density function of the noisy dynamical system is forecasted by recursively computing the transition probability of the internal state by using a Monte Carlo method. This paper is organized as follows; in section \ref{sec:lstm_basic}, the basic structure of LSTM is reviewed. The algorithms to learn the probability density via discretization and to make a forecast of the time evolution of noisy dynamical system are presented in sections \ref{sec:numeric} -- \ref{sec:SMC}. The behaviors of the proposed RNN model is thoroughly studied in section \ref{sec:experiments}. Finally, conclusions are provided in section \ref{sec:conclusions}

\section{Deep Learning algorithm}

In this section, first the basic equations of the Long Short-Term Memory network are reviewed. Then, a numerical discretization procedure to learn the probability distribution of a noisy dynamical system is presented and a regularized cross-entropy loss function is introduced to obtain a penalized maximum likelihood estimator. Finally, a Monte Carlo procedure for a multi-step forecast is outlined. 

\subsection{Review of Long Short-Term Memory network} \label{sec:lstm_basic}

The Long Short-Term Memory network was introduced to consider a delay-time process, where the state of a system at time $t$ is affected by a past event at $t-\tau$ \cite{Hochreiter97}. The basic equations of LSTM unit proposed by \cite{Gers00} consist of a set of nonlinear transformations of an input variable $z \in \mathbb{R}^m$;
\begin{align}
\text{input gate:}&~~ \bm{G}_i = \bm{\varphi}_S \circ \mathcal{L}^{N_c}_i (\bm{z} ), \label{eqn:in_gate}\\
\text{forget gate:}&~~ \bm{G}_f =  \bm{\varphi}_S \circ \mathcal{L}^{N_c}_f (\bm{z} ), \label{eqn:forget_gate}\\
\text{output gate:}&~~ \bm{G}_o =  \bm{\varphi}_S \circ \mathcal{L}^{N_c}_o (\bm{z} ),\label{eqn:out_gate}\\
\text{internal state:}&~~ \bm{s}_{t} = \bm{G}_f \odot \bm{s}_{t-1} + \bm{G}_i \odot \left( \bm{\varphi}_T\circ \mathcal{L}^{N_c}_z (\bm{z}) \right), \label{eqn:cell_state}\\
\text{output:}&~~ \bm{h}_{t} = \bm{G}_o \odot \bm{\varphi}_T(\bm{s}_t), \label{eqn:cell_out}
\end{align}
in which $\bm{\varphi}_S$ and $\bm{\varphi}_T$, respectively, denote the sigmoid and hyperbolic tangent functions, $\mathcal{L}^n$ is a linear transformation operator, $N_c$ is the number of the LSTM units, $s_t$ and $h_t$ represent, respectively, the internal state and the output of the LSTM network, and $\bm{a} \odot \bm{b}$ denotes a component-wise multiplication of two vectors. The linear transformation operator is defined as
\[
\mathcal{L}^{n}(\bm{x}) = \bm{W}\bm{x}+\bm{B},
\]
where $\bm{W}\in \mathbb{R}^{n \times m}$ for $\bm{x}\in\mathbb{R}^m$ and $\bm{B}\in\mathbb{R}^n$ denote a weight matrix and a bias vector, respectively.

\begin{figure}
  \centering
  \includegraphics[height=5cm]{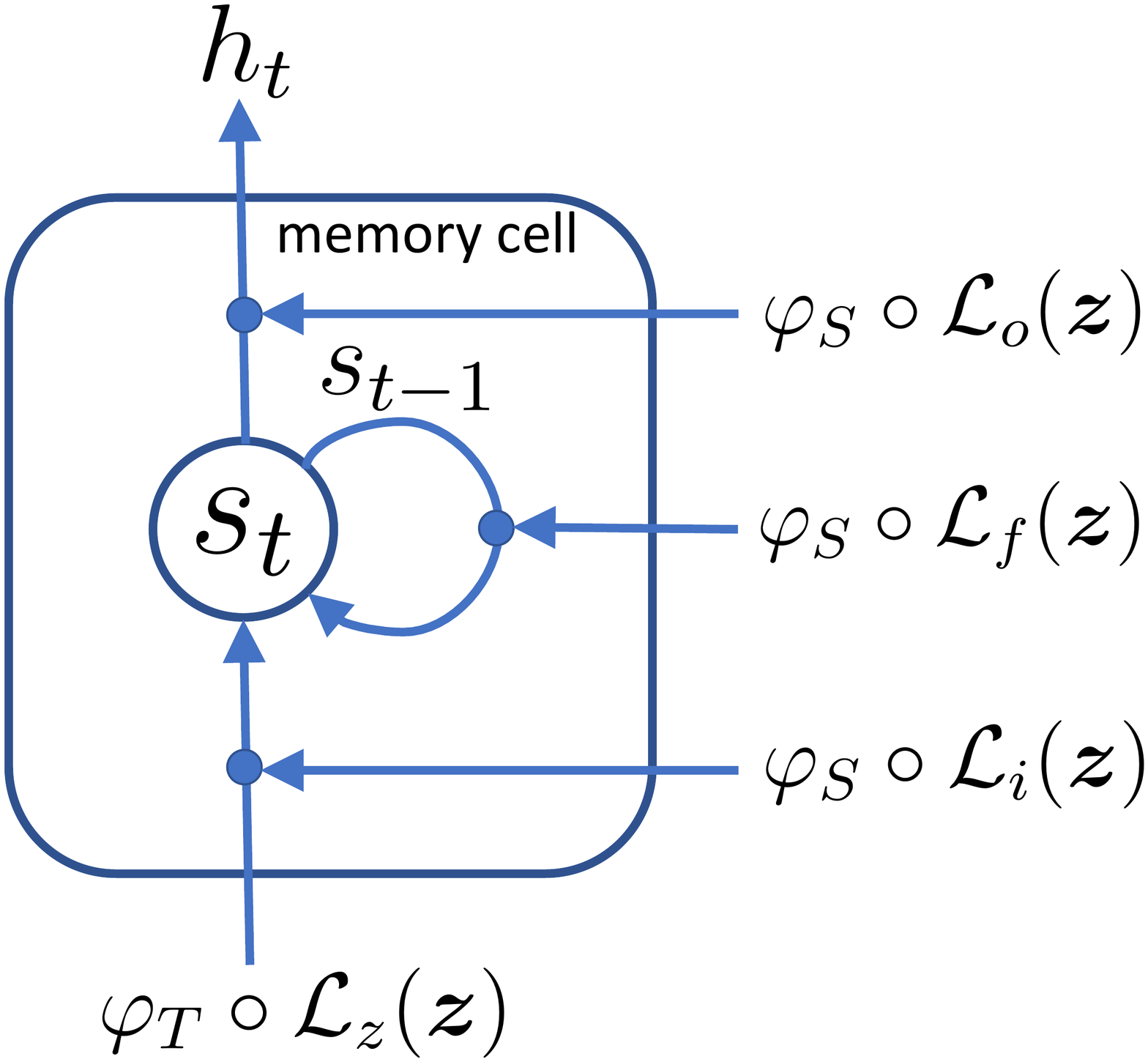}\hspace{0.5cm}
  \includegraphics[height=5cm]{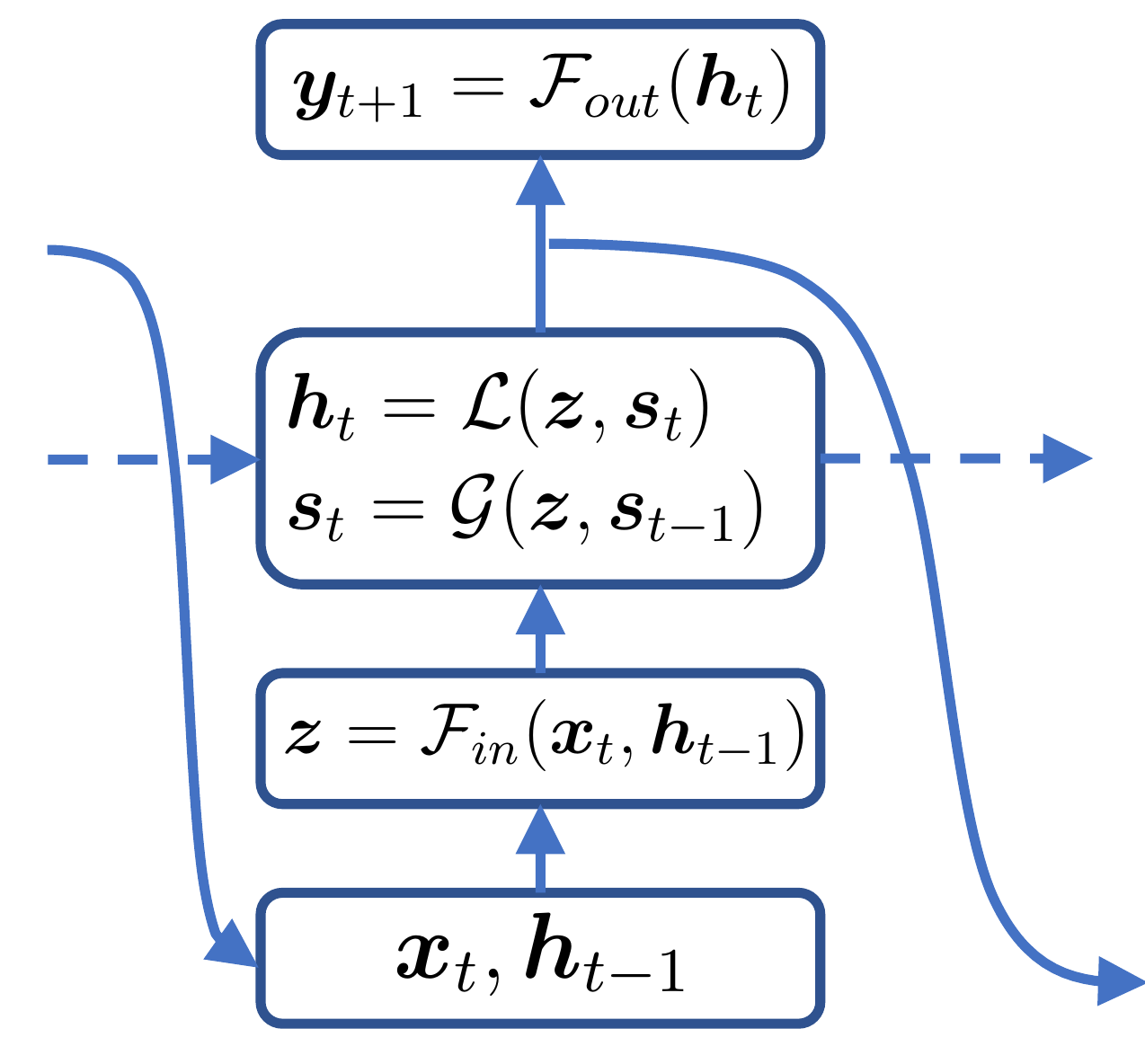}
  \caption{(a) A sketch of Long Short-Term Memory unit and (b) a typical architecture of LSTM network.} \label{fig:LSTM_basic}
\end{figure}

Figure \ref{fig:LSTM_basic} (a) shows a sketch of one LSTM unit. The sigmoid function, which varies between zero and one, is employed in the gate functions. An LSTM network learns from the data when to close ($\varphi_S(x) = 0$) or open ($\varphi_S(x) = 1$) each gates to control information flow into and out of the LSTM units. Equation \ref{eqn:cell_state} shows that, when the input gate is closed, $\bm{G}_i = 0$, and the forget gate is inactive, $\bm{G}_f = 1$, the internal state of the LSTM unit is conserved. Hence, the information stored in an LSTM unit can be carried for a long time period. 

Figure \ref{fig:LSTM_basic} (b) outlines a typical LSTM network architecture to make a prediction of a target variable at the next time step, $\bm{y}_{t+1}$, from an input variable at the current time, $\bm{x}_t$. An LSTM network is ``fully connected'', meaning that the output of LSTM in the previous time step, $\bm{h}_{t-1}$, is used as an input to the LSTM itself. The updated internal state, $\bm{s}_t$, is carried to the next step (dashed arrow). There are two additional feedforward neural networks, $\mathcal{F}_{in}$ and $\mathcal{F}_{out}$, before and after the LSTM, which connects the input variable to the LSTM network and performs a nonlinear transformation of the output of the LSTM network to compute the target variable.


LSTM can be understood as a nonlinear state-space model. 
{
Observe that, from the anti-symmetry of $\varphi_S$, $\bm{G}_f(\bm{z}) = 1-\bm{G}_f(-\bm{z})$, and the last term in (\ref{eqn:cell_state}) depends only on $\bm{z}$, $\bm{G}_i(\bm{z}) \odot \left( \bm{\varphi}_T\circ \mathcal{L}^{N_c}_z (\bm{z}) \right)=g^*(\bm{z})$.
Then, the update rule of $\bm{s}$ in (\ref{eqn:cell_state}) can be written as}
\begin{equation}
s^{(i)}_{t+1} = \left[ 1 - f^{(i)}(\bm{z}_t) \delta t \right] s^{(i)}_t + g^{(i)}(\bm{z}_t) \delta t,~~\text{for}~i=1,\cdots,N_c,
\end{equation}
in which $\delta t$ is the sampling interval of the data, and $ 0 < f(\cdot) < \delta t^{-1}$ and $|g(\cdot)| < \delta t^{-1}$ are some unknown functions. The time evolution of the internal state is essentially a forward Euler scheme for a system of relaxation processes,
\begin{equation} \label{eqn:dyn_phase}
\frac{d s^{(i)}(t) }{d t} = - f^{(i)}(\bm{z}) s^{(i)}(t) + g^{(i)}(\bm{z}).
\end{equation}
The artificial neural network operating on $\bm{x}_t$ and $\bm{h}_{t-1}$ provides nonlinear estimations of the relaxation ($f$) and forcing ($g$) functions. Once LSTM computes the internal phase dynamics by solving (\ref{eqn:dyn_phase}), another artificial neural network, $\mathcal{F}_{out}$ in figure \ref{fig:LSTM_basic} (b), is used for a nonlinear projection of the internal dynamics $\bm{s}_t$ onto the phase space of the target variable, $\bm{y}_{t+1}$. 

It is important to note that the time evolution of $\bm{s}$ is given as  a set of relaxation equations. For a simulation by LSTM, we need to specify the initial conditions, i.e., $\bm{s}_0$ and $\bm{h}_0$. It is challenging, if not impossible, to find a correct initial condition for the data-driven simulation. But, equation (\ref{eqn:dyn_phase}) suggests that the effects of the initial conditions will eventually vanish after some spin-up time and the system becomes stationary.

\subsection{LSTM for noisy dynamical system} \label{sec:numeric}

Here, we consider noisy time series data from a dynamical system. Let $\yhat(t)$ be a noisy observation of a dynamical system,
\begin{equation} \label{eqn:dyn_generic}
  \frac{\partial y}{\partial t} = f(y,\bm{u}),
\end{equation}
in which $\bm{u}$ is an exogenous forcing. For simplicity, we consider a univariate dynamical system, but the modeling framework can be easily extended to a multivariate time series. In general, the ground truth, $y(t)$, is not accessible and we observe only a discrete, corrupted time series,
\begin{equation} \label{eqn:noisy}
\yhat_t = y(t) + \epsilon_t  = y_t + \epsilon_t,
\end{equation}
where $\epsilon_t$ is a white noise. Hereafter, a subscript $t$ indicates a projection onto a discrete space, e.g., $y_{t+n} = \int y(s) \delta(t+n\delta t -s ) ds$, and a sampling interval is denoted by $\delta t$. Here, we are interested in a prediction of the probability distribution of $\yhat$, $p(\yhat_{t+1} | \widehat{\bm{Y}}_{0:t},\bm{U}_{0:t})$, given the past trajectories of the observation, $\widehat{\bm{Y}}_{0:t} = (\yhat_0,\cdots,\yhat_t)$, and the exogenous forcing, $\bm{U}_{0:t} = (\bm{u}_0,\cdots,\bm{u}_t)$. 

Note that, in LSTM, the probability distribution of $\yhat$ can be represented as a Markov process, i.e., $p(\yhat_{t+1} | \widehat{\bm{Y}}_{0:t},\bm{U}_{0:t}) = p(\yhat_{t+1}|\bm{h}_t)$. From the LSTM equations (\ref{eqn:in_gate}--\ref{eqn:cell_out}), we can summarize the overall algorithm in the following three steps, 
\begin{align}
\bm{s}_t &= \Psi_s(\bm{h}_{t-1},\bm{s}_{t-1},\bm{x}_t), \label{eqn:st} \\
\bm{h}_t & = \Psi_h(\bm{h}_{t-1},\bm{s}_t,\bm{x}_t), \label{eqn:ht}\\
\yhat_{t+1} &= \Psi_y(\bm{h}_t), \label{eqn:yt}
\end{align}
in which $\bm{x}_t = (\yhat_t,\bm{u}_t)$, $\Psi_h = \bm{s}_t \odot \bm{G}_o$, and $\Psi_y$ is a feed forward neural network, shown as $\mathcal{F}_{out}$ in figure \ref{fig:LSTM_basic} (b). It is clear that, once $\bm{h}_t$ is given, $\yhat_{t+1}$ is obtained independently from the past trajectory. In other words, the past information is stored in the internal dynamical system of LSTM and $\yhat_{t+1}$ becomes conditionally independent from $\widehat{\bm{Y}}_{0:t}$ and $\bm{U}_{0:t}$. Hence, we first consider the last part of the LSTM of estimating the probability distribution of a target variable, $\yhat_{t+1}$, given an input vector $\bm{h}_t$, $p(\yhat_{t+1}|\bm{h}_t)$. 

Here, we use a notation $p(y|\bm{x})$, instead of $p(\yhat_{t+1}|\bm{h}_t)$, for simplicity. Suppose there is a mapping $\mathcal{C}:\mathbb{R} \rightarrow \mathbb{N}_+$, such that
\begin{equation}
\mathcal{C}(y) = k, ~\text{if}~\alpha_k < y \le \alpha_{k+1}.
\end{equation}
Here, $\bm{\alpha} \in \mathbb{R}^{K+1}$ is a set of ordered real numbers; $\alpha_1 < \alpha_2 < \cdots < \alpha_{K+1}$. Then, we can define a discrete probability as
\begin{equation}
P(k|\bm{x}) = \int_{\alpha_k}^{\alpha_{k+1}} p(y|\bm{x}) dy,~~\text{for}~~k=1,\cdots,K.
\end{equation}
The discrete probability, $P(k|\bm{x})$, is a numerical discretization of the continuous probability distribution, $p(y|\bm{x})$. In the computer science literature, $P(k|\bm{x})$ is called the ``class probability'', as it indicates the probability of $y$ belonging to the $k$-th class, or interval, $\mathcal{I}_k = (\alpha_k,\alpha_{k+1})$. After the discretization, the original problem of estimating a continuous probability function is converted into a multi-label classification problem. 

Suppose there is a data set $\bm{D} = \{(y_i,\bm{x}_i); i = 1,\cdots, N\}$. Then, we can create a new data set by applying $\mathcal{C}$, such that $\bm{D}_C = \{(c_i,\bm{x}_i);  c_i = \mathcal{C}(y_i)~ \text{and}~  i = 1,\cdots, N\}$. The likelihood function for $\bm{D}_C$ is the generalized Bernoulli distribution, or multinoulli distribution, \cite{Bishop06} 
\begin{equation}
P(\bm{c}|\bm{X};\bm{\theta}) = \prod_{n=1}^N \prod_{k=1}^K P(k|\bm{x}_n;\bm{\theta})^{\delta_{c_n k}}.
\end{equation}
Here, $\bm{X} = ( \bm{x}_1,\cdots,\bm{x}_n )$, $\bm{\theta}$ is the parameters of the artificial neural network, and $\delta_{ij}$ is the Kronecker delta function. The discrete probability, $P(k|\bm{x})$ is modeled by an artificial neural network, $\bm{\Psi}(\bm{x};\bm{\theta})$, with a softmax function. The output of the artificial neural network, $\bm{\Psi}(\bm{x};\bm{\theta})$, is a vector of length $K$, and the softmax function is defined as,
\begin{equation}
P(k|\bm{x};\bm{\theta}) = P_k(\bm{\Psi}) = \frac{\exp(\Psi_k)}{\sum_{l=1}^K \exp(\Psi_l)}.
\end{equation}
By definition, the softmax function is $P_k > 0$ and $\sum_k P_k = 1$. 

The parameters of the artificial neural network, $\bm{\theta}$, are estimated by a maximum likelihood method, or by minimizing a negative log likelihood. Let $l(\bm{\theta})$ be a negative log likelihood function,
\begin{equation} \label{eqn:NLL}
l(\bm{\theta}) = - \log P(\bm{c}|\bm{X};\bm{\theta}) =  - \sum_{n=1}^N \sum_{k=1}^K \delta_{c_n k} \log P_k^n(\bm{\Psi}).
\end{equation}
Here, $P_k^n(\bm{\Psi})$ is a short notation of $P_k(\bm{\Psi}(\bm{x}_n;\bm{\theta}))$. A maximum likelihood estimator is
\begin{equation} \label{eqn:ce_opt}
\widehat{\bm{\theta}} = \underset{\bm{\theta} \in \mathbb{R}^{N_w} }{\text{arg min}}  \left\{ - \sum_{n=1}^N \sum_{k=1}^K \delta_{c_n k} \log P_k(\bm{\Psi}(\bm{x}_n;\bm{\theta})) \right\},
\end{equation}
in which $N_w$ is the total number of parameters in $\Psi(\bm{x};\bm{\theta})$. In deep learning, the minimization problem is usually solved by a gradient descent method,
\begin{equation} \label{eqn:ce_update}
\widehat{\bm{\theta}}^{k+1} = \widehat{\bm{\theta}}^k - \eta \bm{\nabla}_{\bm{\theta}} l(\bm{\theta})|_{ \widehat{\bm{\theta}}^k},
\end{equation}
where the superscript $k$ is the number of iteration and $\eta$ is a learning rate. The gradient of $l(\bm{\theta})$ with respect to the output vector of $\Psi(\bm{x};\bm{\theta})$ is
\begin{equation} \label{eqn:dl_da}
\frac{\partial l(\bm{\theta})}{\partial \Psi_i} = \sum_{n=1}^N \sum_{j=1}^K\frac{\partial l(\bm{\theta})}{\partial P^n_j(\bm{\Psi})}\frac{\partial P^n_j(\bm{\Psi})}{\partial \Psi_i} = \sum_{n=1}^N \left( P^n_i - \delta_{c_n i} \right),
\end{equation}
and the full gradient is simply
\begin{equation} \label{eqn:full_grad}
\bm{\nabla}_{\bm{\theta}} l(\bm{\theta}) = \sum_{i=1}^K \frac{\partial l(\bm{\theta})}{\partial \Psi_i}\frac{\partial \Psi_i}{\partial \bm{\theta}}.
\end{equation}
The gradient of the output vector with respect to the parameters, $\partial \Psi_i / \partial \bm{\theta}$, in (\ref{eqn:full_grad}) can be easily computed by a back-propagation algorithm \cite{GoodfellowBengio16}.

The negative log likelihood function (\ref{eqn:NLL}) corresponds to the cross entropy between probability distributions from the data and the model. This kind of cross-entropy minimization is one of the most widely used methods to train a deep neural network for classification. However, the cross-entropy minimization does not explicitly guarantee the smoothness of the estimated distribution. In (\ref{eqn:ce_opt}), the multinoulli loss function depends only on $\bm{P}(\bm{\Psi})$ of a correct label, $\delta_{c_n k}$. As a result, equation (\ref{eqn:dl_da}) indicates that, in the gradient decent steps (\ref{eqn:ce_update}), every $P_i(\bm{\Psi})$ except for the one for the correct label, $P_{c_n}(\bm{\Psi})$, is penalized in the same way. So, there is no explicit way to guarantee smoothness of the estimated distribution. In the conventional classification tasks, a geometric proximity between the classes is not relevant. In the present study, however, the softmax function is used as a discrete approximation to a probability density function, which requires a smoothness in $\bm{P}(\bm{\Psi})$. 

To impose the smoothness constraint, we propose a penalized maximum likelihood method, inspired by a non-parametric density estimation problem \cite{Silverman86}. The penalized maximum likelihood estimator is computed by adding a regularization to the standard cross-entropy loss function;
\begin{equation} \label{eqn:reg_ce_opt}
\bm{\theta}^* = \underset{\bm{\theta} \in \mathbb{R}^{N_w} }{\text{arg min}}   \sum_{n=1}^N \left\{ \sum_{k=1}^K -\delta_{c_nk} \log P_k^n + \lambda \left( \bm{L} \bm{P}^n \right)^T \bm{D} \left(\bm{L}\bm{P}^n\right) \right\},
\end{equation}
in which $\lambda$ is a penalty parameter, $\bm{D}$ is a diagonal matrix, $D_{ij} = \Delta_{i+1}\delta_{ij}$ for $i = 1,\cdots,K-2$, and $\Delta_i  = |\mathcal{I}_i| = \alpha_{i+1}-\alpha_i$. The Laplacian matrix $\bm{L}\in \mathbb{R}^{K-2,K}$ is
\begin{equation}
\bm{L} = 
\begin{bmatrix}
l_{1a}&l_{1b}&l_{1c}&0&\cdots&0\\
0&l_{2a}&l_{2b}&l_{2c} &\cdots&0\\
\hdotsfor{6}\\
0&\cdots&0&l_{(K-2)a}&l_{(K-2)b}&l_{(K-2)c}
\end{bmatrix}.
\end{equation}
Here,
\begin{equation}
l_{ia} = \frac{2}{\delta^-_i(\delta^-_i-\delta^+_i)}\frac{1}{\Delta_i},~l_{ib} = \frac{2}{\delta^-_i\delta^+_i}\frac{1}{\Delta_{i+1}},~l_{ic} = \frac{2}{\delta^+_i(\delta^+_i-\delta^-_i)}\frac{1}{\Delta_{i+2}},
\end{equation}
and
\begin{equation}
\delta^-_i = -\frac{1}{2}(\Delta_{i+1} + \Delta_i),~\delta^+_i = \frac{1}{2}(\Delta_{i+2}+\Delta_{i+1}).
\end{equation}
Suppose there is a smooth function, $g(y)$, such that
\[
g(\alpha_{i+1/2}) = P_i,
\]
where $\alpha_{i+1/2} = 0.5(\alpha_i+\alpha_{i+1})$. The regularization in (\ref{eqn:reg_ce_opt}) corresponds to
\begin{equation}
  \left( \bm{L} \bm{P} \right)^T \bm{D} \left(\bm{L}\bm{P} \right) \simeq \int (g''(y))^2 dy.
\end{equation}
This regularized cross-entropy (RCE) aims to smooth out the estimated distribution by penalizing local minima or maxima. 

Let $l^*(\bm{\theta})$ denote RCE (\ref{eqn:reg_ce_opt}), the gradient of $l^*(\bm{\theta})$ with respect to $\bm{\Psi}(\bm{x}^n;\bm{\theta})$ is
\begin{equation}\label{eqn:dl_da_rce}
\frac{\partial l^*(\bm{\theta})}{\partial \Psi_i} = \sum_{n=1}^N \left[ P^n_i \left\{  1 + 2 \lambda \left(  \bm{M}_{i,\cdot}\bm{P}^n - {\bm{P}^n}^T\bm{M}\bm{P}^n \right)    \right\}  - \delta_{c_n i}  \right],
\end{equation}
in which $\bm{M}=\bm{L}^T\bm{D}\bm{L}$ and $\bm{M}_{i,\cdot} \bm{P} = \sum_j M_{ij}P_j$. Comparing to (\ref{eqn:dl_da}), the additional computational cost to consider the smoothness is only the maxtrix-vector multiplications in the curly bracket in (\ref{eqn:dl_da_rce}).

Extending the RCE minimization to a time series data is straightforward. Suppose we have a data set, which consist of $N$ time series with length $T$; $\mathcal{D}_y = \{(\bm{\yhat}^n,\bm{u}^n); \bm{\yhat}^n \in \mathbb{R}^T~\text{and}~\bm{u}^n \in \mathbb{R}^{T \times m}~\text{for}~n =1,\cdots,N\}$. By applying $\mathcal{C}$, the data set can be converted to $\mathcal{D}_c = \{(\bm{c}^n,\bm{\yhat}^n,\bm{u}^n); \bm{c}^n =\mathcal{C}(\bm{\yhat}^n), \bm{\yhat}^n \in \mathbb{R}^T,~\text{and}~\bm{u}^n \in \mathbb{R}^{T\times m}~\text{for}~n =1,\cdots,N\}$. The data likelihood function for $\mathcal{D}_c$ is given as
\begin{equation}
P(\bm{C}|\bm{X};\bm{\theta}) =  \prod_{n=1}^N \prod_{l=1}^{T-1} \prod_{k=1}^K P(k|\bm{x}^n_l,\bm{h}^n_{(l-1)},\bm{s}^n_{(l-1)};\bm{\theta})^{\delta_{c^n_{l+1} k}},
\end{equation}
where $\bm{x}^n_l = (\yhat^n_l,\bm{u}^n_l)$. 
The regularized cross-entropy for LSTM is
\begin{equation}
L(\bm{\theta}) = \sum_{l=1}^{T-1} l_l^*(\bm{\theta}),
\end{equation}
and,
\begin{equation}
l_l^*(\bm{\theta}) = \sum_{n=1}^N \left\{ \sum_{k=1}^K -\delta_{c^n_{l+1} k} \log P^n_{{l+1}_k} + \lambda {\bm{P}^n_{l+1}}^T \bm{M}\bm{P}^n_{l+1} \right\},
\end{equation}
in which $\bm{P}^n_{l+1} = \bm{\Psi}_y(\bm{h}^n_l)$. The regularized cross-entropy minimization problem for LSTM can be solved by the standard back-propagation through time algorithm (BPTT) \cite{GoodfellowBengio16}. In BPTT, the temporal structure of error is propagated backward in time, e.g.,
 \begin{equation}
 \bm{\nabla}_{\bm{h}_t} L(\bm{\theta}) = \frac{\partial L(\bm{\theta})}{\partial \bm{h}_t} + \frac{\partial L(\bm{\theta})}{\partial \bm{h}_{t+1}}\frac{\partial \bm{h}_{t+1}}{\partial \bm{h}_t} =  \frac{\partial l^*_t(\bm{\theta})}{\partial \bm{h}_t} + \frac{\partial l^*_{t+1}(\bm{\theta})}{\partial \bm{h}_{t+1}}\frac{\partial \bm{h}_{t+1}}{\partial \bm{h}_t} .
 \end{equation}
It is explicitly shown that the gradient of the loss function at $t$ is linked to the loss function at the next time step, $l^*_{t+1}$.  For more details, see \cite{Gers01}.
 
\subsection{Monte Carlo method for multiple-step forecast} \label{sec:SMC}

One of the major interests in time series modeling is to make a multiple-step forecast of the state of the system conditioned on the past observations. For a noisy dynamical system with an exogenous forcing, (\ref{eqn:dyn_generic} -- \ref{eqn:noisy}), a multi-step forecast is to compute the future probability distribution, $p(\yhat_{t+n}|\widehat{\bm{Y}}_{0:t},\bm{U}_{0:t+n-1})$ for $n > 1$, given the past trajectories, $(\widehat{\bm{Y}}_{0:t},\bm{U}_{0:t})$, and a future forcing scenario, $\bm{U}_{t+1:t+n-1}$. 
In other words, we are interested in computing a temporal evolution of the probability distribution for a given forcing scenario.

A multiple-step forecast can be achieved by successively applying the deterministic transformations in (\ref{eqn:st} -- \ref{eqn:yt}). Suppose the data is given up to time $t$, i.e., we have $(\widehat{\bm{Y}}_{0:t},\bm{U}_{0:t})$. The probability distribution at $t+1$ is computed by a deterministic update,
\begin{equation}\label{eqn:update_1}
\begin{cases}
\bm{s}_{t} &= \Psi_s(\bm{h}_{t-1},\bm{s}_{t-1},\yhat_t,\bm{u}_t),\\
\bm{h}_t &= \Psi_h(\bm{h}_{t-1},\bm{s}_t,\yhat_t,\bm{u}_t), \\
p(\yhat_{t+1}|\bm{h}_t) &= \Psi_y(\bm{h}_t).
\end{cases}
\end{equation}
Again, from the state-space model argument in sections \ref{sec:lstm_basic} and \ref{sec:numeric}, we have $p(\yhat_{t+1} | \widehat{\bm{Y}}_{0:t},\bm{U}_{0:t}) = p(\yhat_{t+1}|\bm{h}_t)$.
In the next time step at $t+2$, the internal state is updated as,
\[
\bm{s}_{t+1} = \Psi_s(\bm{h}_{t},\bm{s}_{t},\yhat_{t+1},\bm{u}_{t+1}).
\]
{
Because the observation is available only up to $\yhat_t$, $\yhat_{t+1}$ becomes a random variable, of which distribution is computed in (\ref{eqn:update_1}).} Hence, $\bm{s}_{t+1}$ becomes a random variable and the probability distribution of $\bm{s}_{t+1}$ is fully determined by $(\bm{h}_{t},\bm{s}_{t},\bm{u}_{t+1})$, i.e., $p(\bm{s}_{t+1}|\bm{s}_{t},\bm{h}_{t},\bm{u}_{t+1})$. Similarly, the probability distribution of $\bm{h}_{t+1}$ is given as $p(\bm{h}_{t+1}|\bm{s}_{t+1},\bm{h}_t,\bm{u}_{t+1})$. Finally, the probability distribution of $\yhat$ at $t+2$, $p(\yhat_{t+2}|\widehat{\bm{Y}}_{0:t},\bm{U}_{0:t+1})$, is 
\begin{equation}
p(\yhat_{t+2}|\widehat{\bm{Y}}_{0:t},\bm{U}_{0:t+1}) = \iint p(\yhat_{t+2}|\bm{h}_{t+1})p(\bm{h}_{t+1},\bm{s}_{t+1}|\bm{h}_t,\bm{s}_{t},\bm{u}_{t+1}) d\bm{s}_{t+1} d\bm{h}_{t+1}.
\end{equation}
Here, a product rule is used,
\[
p(\bm{h}_{t+1},\bm{s}_{t+1}|\bm{h}_{t},\bm{s}_t,\bm{u}_{t+1}) = p(\bm{h}_{t+1}|\bm{s}_{t+1},\bm{h}_{t},\bm{u}_{t+1}) p(\bm{s}_{t+1}|\bm{s}_{t},\bm{h}_{t},\bm{u}_{t+1}).
\]
From a recurrent relation, the multiple-step forecast is
\begin{align} \label{eqn:pred_prob}
&p(\yhat_{t+n}|\widehat{\bm{Y}}_{0:t},\bm{U}_{0:t+n-1})  = \nonumber \\
&\idotsint p(\yhat_{t+n}|\bm{h}_{t+n-1}) \prod_{i=1}^{n-1}p(\bm{H}_{t+i}|\bm{H}_{t+i-1},\bm{u}_{t+i})d\bm{H}_{t+i},~\text{for}~n > 1,
\end{align}
in which $\bm{H}_t = (\bm{h}_t,\bm{s}_t)$. Equation (\ref{eqn:pred_prob}) is computed by using (\ref{eqn:st} -- \ref{eqn:ht}) as an initial condition. In practice, directly computing the probability distribution of the LSTM states, $p(\bm{H}_{t+1}|\bm{H}_{t})$, is intractable because of the high dimensionality, $\bm{H} \in \mathbb{R}^{2N_c}$. Here, we outline a Monte Carlo simulation to approximate the high-dimensional integration in (\ref{eqn:pred_prob}) in Algorithm \ref{alg:MC}.

\begin{algorithm}[!th]
	\caption{Monte Carlo method for a multi-step forecast}
	\label{alg:MC}
	{
	\hspace*{\algorithmicindent} \textbf{Input}: $\widehat{\bm{Y}}_{0:t}$, $\bm{U}_{0:t+n-1}$, MC sample size ($N_s$), forecast horizon $n$\\
	\hspace*{\algorithmicindent} \textbf{Output}: $p(\yhat_{t+1}|\widehat{\bm{Y}}_{0:t},\bm{U}_{0:t}),\cdots,p(\yhat_{t+n}|\widehat{\bm{Y}}_{0:t},\bm{U}_{0:t+n-1})$ \\
	}
	\begin{algorithmic}
		\STATE Initialize LSTM states: $\bm{s}_0 = \bm{h}_0 = \bm{0}$			
		\STATE	
		\STATE A sequential update of LSTM up to time $t$ using the data, { $\bm{x}_t = (\yhat_t,\bm{u}_t)$.} 
		\FOR{ $j=1,t$}
			\STATE $\bm{s}_j = \Psi_s(\bm{s}_{j-1},\bm{h}_{j-1},\bm{x}_j)$
			\STATE $\bm{h}_j = \Psi_h(\bm{s}_j,\bm{h}_{j-1},\bm{x}_j)$
		\ENDFOR
		
		\STATE
		\STATE  Make $N_s$ replicas of the internal states 
		\[
		\bm{s}^{(1)}_t=\cdots=\bm{s}^{(N_s)}_t=\bm{s}_t,~~\bm{h}^{(1)}_t=\cdots=\bm{h}^{(N_s)}_t=\bm{h}_t,.		
		\]
		
		\FOR{ $j=1,n$ } 
			\FOR{$i=1,{N_s}$}
			\STATE Compute the predictive distribution of $\yhat^{(i)}_{t+j}$ for each sample
			\[
			\bm{P}^{(i)}_{t+j} = \bm{\Psi}_y(\bm{h}^{(i)}_{t+j-1})
			\] 
			\STATE Draw $\yhat^{(i)}_{t+j}$ from the computed distribution:
			    \STATE \hspace{1em} 1. Draw the class label from the discrete distribution: $k^{(i)} \sim \bm{P}^{(i)}_{t+j}$
			    \STATE \hspace{1em} 2. Draw $\yhat^{(i)}_{t+j}$ in $\mathcal{I}_{k^{(i)}}$: ${\yhat^{(i)}_{t+j} \sim \mathcal{U}(\mathcal{I}_{k^{(i)}})^\dagger}$
			\STATE
			\STATE Update the internal states of LSTM:
			\[ \bm{s}^{(i)}_{t+j} = \Psi_s(\bm{s}^{(i)}_{t+j-1},\bm{h}^{(i)}_{t+j-1},\bm{u}_{t+j}) \]
			\[ \bm{h}^{(i)}_{t+j} = \Psi_h(\bm{s}^{(i)}_{t+j},\bm{h}^{(i)}_{t+j-1},\bm{u}_{t+j}) \]
			\ENDFOR
		{
		\STATE Compute $p(\yhat_{t+j}|\widehat{\bm{Y}}_{0:t},\bm{U}_{0:t+j-1})$ by a (kernel) density estimation.
		}
		\ENDFOR
		
	{${}^{\dagger}$: $\mathcal{U}(\mathcal{I}_m)$ denotes a uniform distribution in a grid cell, $\mathcal{I}_m = (\alpha_{m-1},\alpha_m)$.}
	\end{algorithmic}
\end{algorithm}

\section{Numerical experiments} \label{sec:experiments}

In this section, numerical experiments of the LSTM simulations of noisy dynamical systems are presented. The  same LSTM architecture is used in all of the numerical experiments. The input to the LSTM network is computed as 
\begin{equation} \label{eqn:f_in_test}
\bm{z}_t = \left( \mathcal{L}^{N_c}_{in,2}\circ \left( \varphi_T \circ \mathcal{L}^{N_c}_{in,1} \right) \right)(\bm{x}_t) + \mathcal{L}^{N_c}_{in,3}(\bm{h}_{t-1}).
\end{equation}
The output network, $\bm{\Psi}_y(\bm{h}_t)$, is
\begin{equation} \label{eqn:decoder_test}
\bm{P}_{t+1} = (\varphi_{SM} \circ (\mathcal{L}^{K}_{out,2} \circ (\varphi_T \circ \mathcal{L}_{out,1}^{N_c})))(\bm{h}_{t}).
\end{equation}
Here, $K$ is the number of classes (discretization intervals), $N_c$ is the number of LSTM units, $\bm{x}_t = (\yhat_t,\bm{u}_t)$, and $\varphi_{SM}$ is the softmax function.
The number of LSTM units is fixed, $N_c = 128$, unless stated otherwise. 

The parameters of the LSTM are the elements of the weight matrices and bias vectors of $\mathcal{L}$ in (\ref{eqn:in_gate}--\ref{eqn:cell_state}) and (\ref{eqn:f_in_test}--\ref{eqn:decoder_test}). The total number of parameters is $O(10^5)$. For example, when $N_c = 128$, $K = 201$, and $m=1$, the total number of parameters is $\dim(\bm{\theta}) = 191,433$. To obtain a penalized maximum likelihood estimator, RCE is minimized by using a minibatch stochastic gradient descent method, called ADAM \cite{Kingma15}, with a minibath size of 20. To learn the dynamics, the length of a time series in the model training, $T$, should larger than the characteristic timescale of the dynamical system. Here, $T=100$ is used.  
Hereafter, we use DE-LSTM (density-estimation LSTM) to refer the LSTM model proposed in the present study. The data input to and output from DE-LSTM are standardized such that $x_i^* = (x_i - E[x_i])/sd(x_i)$, in which $E[x_i]$ and $sd(x_i)$ are the mean and standard deviation of a variable $x_i$ of the ``training data set'', e.g., $\yhat^* = (\yhat - E[\yhat])/sd[\yhat]$. While the computations are performed for the standardized variables, the results shown in this section are rescaled back to the original scale.

In this study, the temporal gradient of a target, $d\yhat_t = \yhat_{t+1}-\yhat_t$, is considered, instead of $\yhat_t$ itself. In other words, in the training phase, an input data to the model is $\bm{x}_t = (\yhat_t,\bm{u}_t)$, while the target variable is $\mathcal{C}(d\yhat_t)$, to learn $p(d\yhat_t)$. If we model $p(\yhat)$, the range of the data, $I(\yhat) = [\yhat_{min},\yhat_{\max}]$, can be so large that the total number of classes, $K$, becomes too large to cover the entire $I(\yhat)$, or we need to make the bin size, $\Delta_i$, large to reduce $K$. However, the range of $d\yhat$ is much smaller than that of $\yhat$, $|I(d\yhat)| < |I(\yhat)|$, and thus we can afford using a high resolution to model the probability distribution. It is trivial to recover $p(\yhat_{t+1}|\widehat{\bm{Y}}_{0:t})$ from $p(d\yhat_t|\widehat{\bm{Y}}_{0:t})$.

\subsection{Ornstein--Uhlenbeck process} \label{sec:OU_process}

First, we consider the Ornstein--Uhlenbeck process, which is represented by the following stochastic differential equation, 
\begin{equation}
dy(t) = - \frac{1}{\tau}y(t) dt + \xi dW,
\end{equation}
in which $\tau$ is a relaxation timescale and $W$ is the Weiner process. The Ornstein--Uhlenbeck process has a closed form solution;
\begin{align}
E[y(t+n\delta t)|y(t)] &= y(t) \exp( -\frac{n\delta t }{\tau} ), \\
Var[y(t+n\delta t)|y(t)] &= \frac{\xi^2 \tau}{2}\left\{1- \exp( -\frac{2n\delta t }{\tau} ) \right\}.
\end{align}
The parameters used in this example are, $\tau = 1$, $\xi = \sqrt{2}$, and $\delta t = 0.1$. Note that, in this example, we do not distinguish the noisy observation, $\yhat_t$, from the ground truth, $y_t$, because $y_t$ itself is a stochastic process.

\begin{figure}
  \centering
  \includegraphics[height=6cm]{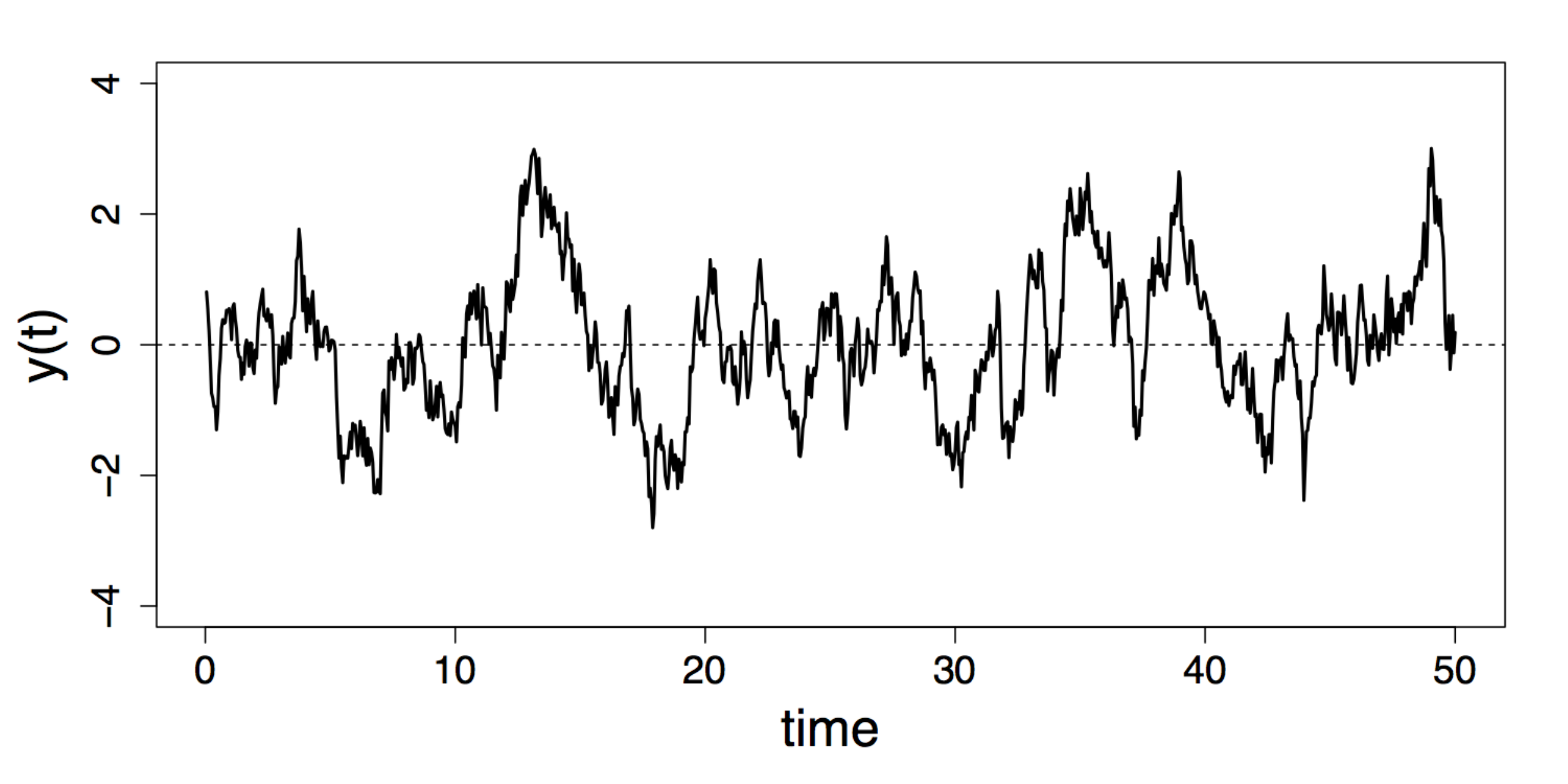}
  \caption{Sample trajectory of Ornstein--Uhlenbeck process.} \label{fig:OU_sample}
\end{figure}

Figure \ref{fig:OU_sample} shows a sample trajectory of the Ornstein-Uhlenbeck process. A simulation is performed for $4\times 10^5 \delta t$ to generate a training set and for another $2 \times 10^3 \delta t$ to make a data set for model validation. The initial learning rate is $\eta_0 = 10^{-3}$ and it is decreased with the number of iteration, $k$, as $\eta_k = \eta_0/(1+10^{-3} k)$. At each iteration, 20 sample trajectories with identical length, $T=100$, are used as a minibatch. The starting point of each sample trajectory is randomly selected from the training data set, $t_0 \in [1,4\times10^5-T]\delta t$. The size of the bins to discretize the probability distribution is uniform, i.e., $\Delta_1 = \cdots = \Delta_K = \delta y$.

\begin{figure}
  \centering
  \includegraphics[height=4.5cm]{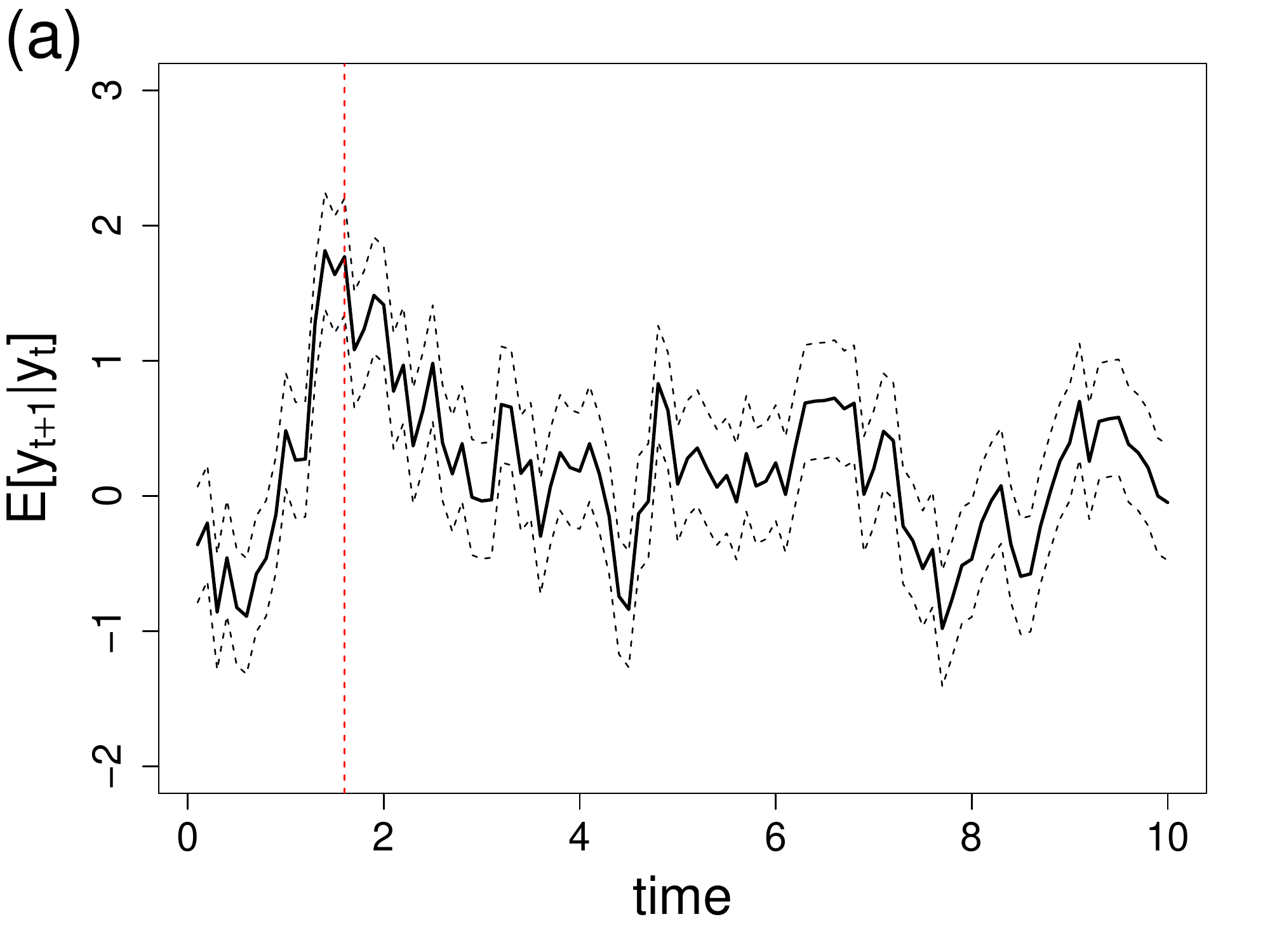}
  \includegraphics[height=4.5cm]{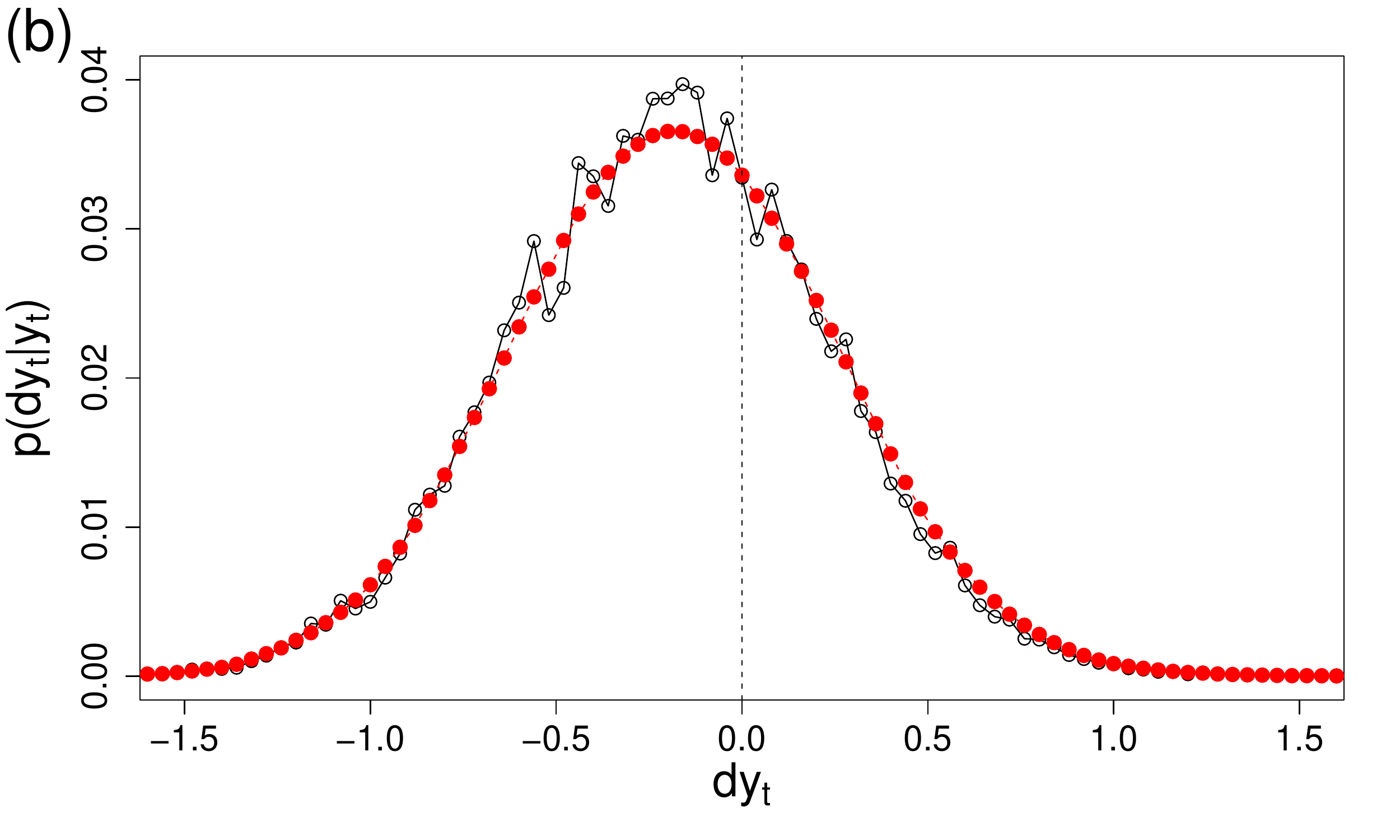}
  \caption{(a) Expectation ($\frac{~~~}{~~~}$) and one-standard-deviation bounds ($\frac{~}{~}\frac{~}{~}\frac{~}{~}$) computed from $p(y_{t+1}|y_t)$ for $\delta y = 0.04$ and $\lambda = 0.1$ and (b) the probability distribution estimated with $\lambda = 0$ ($\circ$) and $\lambda=0.1$ ({\color{red}$\bullet$}).} \label{fig:OU_one_step}
\end{figure}

Figure \ref{fig:OU_one_step} demonstrates an example of the DE-LSTM simulation. In this example, the discretization interval is $\delta y = 0.04$. Here, we consider the next-step prediction problem, in which DE-LSTM is used to compute $p(dy_t|y_t)$. Note that the Ornstein-Uhlenbeck process is a Markov process, $p(dy_t|\bm{Y}_{0:t}) = p(dy_t|y_t)$. The prediction is made as, $E[y_{t+1}|y_t] = y_t + E[dy_t|y_t]$. The expectation and higher-order moments are easily computed by a numerical integration. For example,
\[
E[y] = \sum_{i=1}^K \alpha_{i+1/2} P_i,~~Var[y] =  \sum_{i=1}^K \alpha^2_{i+1/2} P_i - E[y]^2,
\]
in which $\alpha_{i+1/2} = 0.5(\alpha_i + \alpha_{i+1})$. Compared to the deterministic LSTM used for regression, DE-LSTM directly estimates the probability distribution, which provides rich information about the uncertainty in the prediction. For example, as shown in figure \ref{fig:OU_one_step} (a), DE-LSTM can provide the prediction together with uncertainty.

The estimated probability distribution by DE-LSTM is shown in figure \ref{fig:OU_one_step} (b). The probability distribution is computed at the vertical dashed line in figure \ref{fig:OU_one_step} (a). As expected, when the regularization is not used, i.e., $\lambda = 0$, the estimated distribution is bumpy. For $\lambda = 0.1$, it is shown that the estimated distribution becomes smooth.


\begin{table}
\center{
\caption{Normalized root mean-square errors of the expectation.} \label{tbl:OU_mu_error}
\begin{tabular}{c|cccccc}
\hline \hline
\multirow{2}{*}{$\delta y$}& \multicolumn{6}{c}{$\lambda$}\\
 & 0 & $10^{-3}$ & $10^{-2}$ & $10^{-1}$ & $5\times10^{-1}$ & $1$ \\
\hline
0.08 & 1.78 & 0.075 & 0.069 & 0.055 & 0.050 & 0.055\\
0.04 & 0.13 & 0.081 & 0.062 & 0.052 & 0.063 & 0.091\\
\hline \hline
\end{tabular}
}
\end{table}

For a quantitative comparison, a normalized root mean-square error (NRMSE) of $E[y_{t+1}|y_t]$ is computed as
\begin{equation} \label{eqn:def_nrms}
e_\mu = \frac{ \langle (E_{y \sim p_L}[y_{t+1}|y_t] - E_{y\sim p_T}[y_{t+1}|y_t])^2 \rangle^{1/2} }{\langle (E_{y \sim p_T}[y_{t+1}|y_t]-y_t)^2 \rangle^{1/2}},
\end{equation}
in which $\langle \cdot \rangle$ denotes an ensemble average, and $p_T$ and $p_L$ are the true and DE-LSTM probability distributions, respectively. The normalized root mean-square error compares the prediction error between DE-LSTM and a zeroth order prediction, which assumes $y_{t+1} = y_t$. Similarly, NRMSE of the standard deviation is defined as
\begin{equation}
e_{sd} = \frac{sd_{y \sim p_L}[dy]}{sd_{y \sim p_T}[dy]} -1.
\end{equation}
Tables \ref{tbl:OU_mu_error} and \ref{tbl:OU_sd_error} show $e_\mu$ and $e_{sd}$, respectively. 

In Table \ref{tbl:OU_mu_error}, it is shown that DE-LSTM trained with RCE, i.e., $\lambda > 0$, makes a better prediction of the expectation. As $\lambda$ is increased, $e_\mu$ reduces at first, then above a threshold $e_\mu$ starts to increase. Such dependence on the penalty parameter is typical for a penalized maximum likelihood method. It should be noted that $e_\mu$ is not sensitive to the changes in $\lambda$. For example, for $\delta y = 0.04$, the difference between the maximum and minimum $e_\mu$ is only 0.011, when there is a fiftyfold increase in $\lambda$; $\lambda = 0.01$ -- 0.5.

\begin{table}
\center{
\caption{Normalized root mean-square errors of the standard deviation.} \label{tbl:OU_sd_error}
\begin{tabular}{c|cccccc}
\hline \hline
\multirow{2}{*}{$\delta y$}& \multicolumn{6}{c}{$\lambda$}\\
 & 0 & $10^{-3}$ & $10^{-2}$ & $10^{-1}$ & $5\times10^{-1}$ & $1$ \\
\hline
0.08 & 0.0298 & 0.0039 & 0.0054 & 0.0220 & 0.071 & 0.114\\
0.04 & 0.0047 & 0.0030 & 0.0033 & 0.0082 & 0.026 & 0.045\\
\hline \hline
\end{tabular}
}
\end{table}

 It is observed that, when RCE is used, the grid resolution, $\delta y$, does not have a noticeable effect in the estimation of the expectation. Except for $\lambda = 0$, $e_\mu$ for $\delta y = 0.08$ is very close to that of $\delta y = 0.04$. On the other hand, $e_{sd}$ in Table \ref{tbl:OU_sd_error} clearly shows the impact of $\delta y$ on the estimation of the probability distribution. When a fine resolution ($\delta y = 0.04$) is used, not only $e_{sd}$ is smaller than that of $\delta y = 0.08$, but also the sensitivity of $e_{sd}$ to $\lambda$ is much smaller. As $\lambda$ changes from 0.001 to 0.1, $e_{sd}$ for $\delta y = 0.08$ increases from 0.0039 to 0.022, while, for $\delta y = 0.04$, $e_{sd}$ changes only from 0.0030 to 0.0082.

\begin{table}
\center{
\caption{Scaled Kullback-Leibler divergence: $D_{KL}(Q||P)\times10^4$.} \label{tbl:OU_KL}
\begin{tabular}{c|cccccc}
\hline \hline
\multirow{2}{*}{$\delta y$}& \multicolumn{6}{c}{$\lambda$}\\
 & 0 & $10^{-3}$ & $10^{-2}$ & $10^{-1}$ & $5\times10^{-1}$ & $1$ \\
\hline
0.08 & $69.38$ & $0.68$ & $0.31$ & $0.75$ & $5.56$ & $12.29$\\
0.04 & $1.46$ & $0.32$ & $0.16$ & $0.12$ & $0.54$ & $1.34$\\
\hline \hline
\end{tabular}
}
\end{table}

While $e_\mu$ and $e_{sd}$ provide useful information on the behaviors of DE-LSTM, those metrics only compare the first and second moments of the probability distribution. To make a more thorough comparison of the estimated probability distribution, the Kullback--Leibler divergence is computed. Here, the Kullback--Leibler divergence is defined as
\begin{equation}
D_{KL}(Q||P) = \Big\langle -\sum_{i=1}^K \left( Q_i \log \frac{P_i}{Q_i} \right)\delta y \Big\rangle.
\end{equation}
Here, $P_i$ is the $i$-th output of DE-LSTM and $Q_i$ is the true probability distribution,
 \[
 Q_i = \frac{1}{\sqrt{2\pi}\sigma} \int_{\alpha_i}^{\alpha_{i+1}} \exp \left( -\frac{(y-\mu)^2}{2\sigma^2} \right) dy,
 \]
 in which $\mu = E[y_{t+1}|y_t]$ and $\sigma = sd[dy]$. The Kullback--Leibler divergence is non-negative, $D_{KL}(Q||P) \ge 0$ and measures the dissimilarity between two probability distributions, $P$ and $Q$ \cite{Bishop06}. Table \ref{tbl:OU_KL} shows $D_{KL}(Q||P)$ as a function of $\delta y$ and $\lambda$. It is shown that using a fine resolution, i.e., smaller $\delta y$, provides a better approximation of the true probability distribution. At the same time, the estimated probability distribution becomes much less sensitive to the penalty parameter, $\lambda$, at the finer resolution. For $\delta y = 0.04$, there is about fourfold increase in $D_{KL}(Q||P)$ when $\lambda$ changes from $0.001$ to 1. On the other hand, for $\delta y = 0.08$, $D_{KL}(Q||P)$ for $\lambda = 1$ is about 18 times larger than $D_{KL}(Q||P)$ at $\lambda = 0.001$.

\begin{figure}
  \centering
  \includegraphics[height=4.5cm]{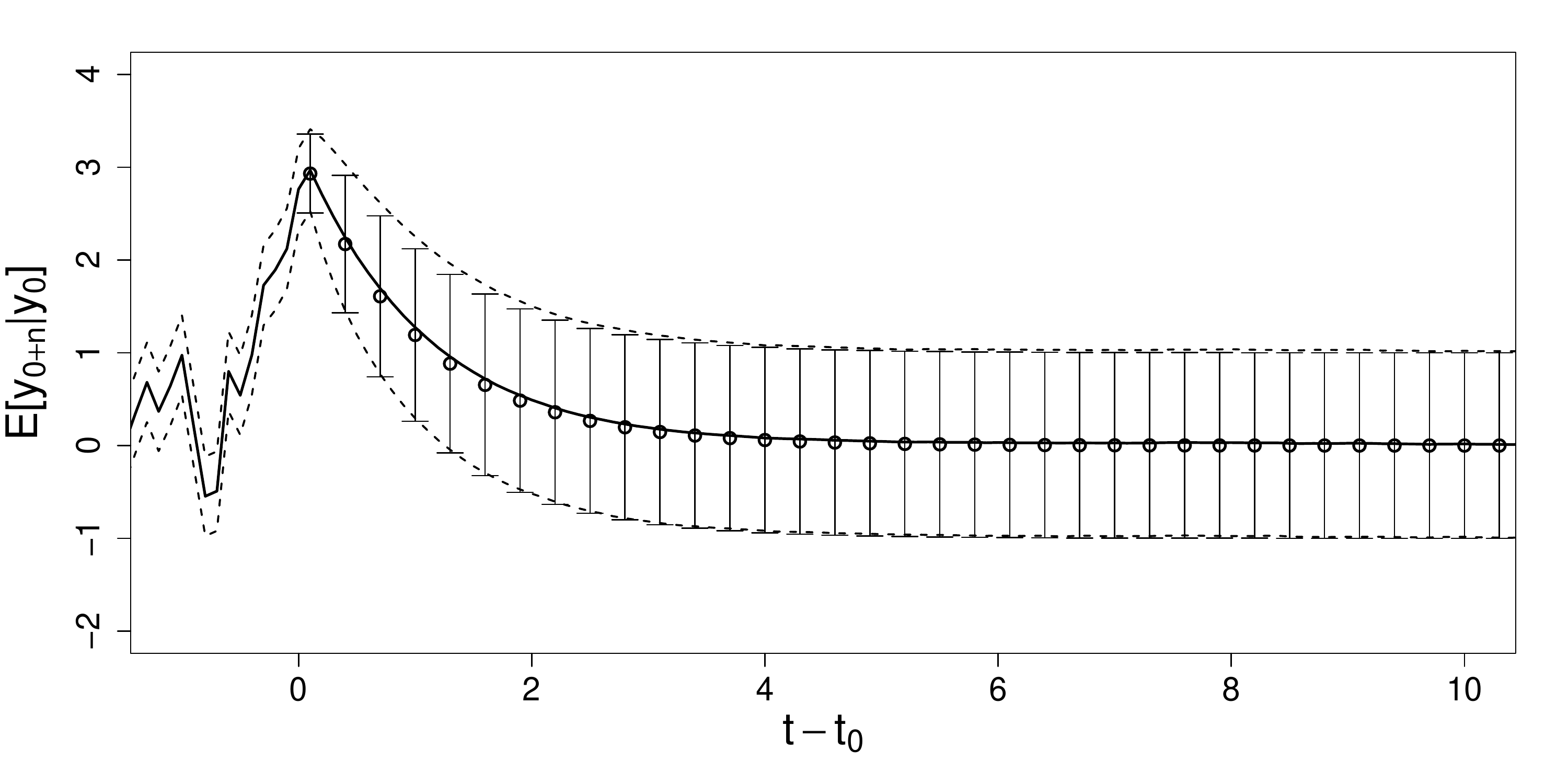}
  \caption{Multiple-step forecast with $\delta y = 0.04$, and $\lambda = 0.1$. The sold line is $E[y_{t_0+n}|y_{t_0}]$ and the dashed lines denote one standard deviation from the expectation. The hollow circles and error bars are the analytical solution. Here, $t_0$ denotes the time of the last data.} \label{fig:OU_multi_step}
\end{figure}

In figure \ref{fig:OU_multi_step}, a multiple-step prediction of DE-LSTM is compared with the analytical solutions. The DE-LSTM is trained with $\delta y = 0.04$ and $\lambda = 0.1$. The sample size of the Monte Carlo method is $50,000$. The data from the Ornstein-Uhlenbeck process is supplied to DE-LSTM for the first 50 time steps and the multiple-step prediction is performed for the next 150 time steps. It is shown that the multiple-step predictions from DE-LSTM agree very well with the analytical solutions of Ornstein-Uhlenbeck process. Here, we define an integral error,
\begin{equation}
e_\mu = \left\{ \frac{ \sum_{i=1}^{150} ( E_{y \sim p_L}[y_{t_0+i}|y_{t_0}] - E_{y\sim p_T}[y_{t_0+i}|y_{t_0}] )^2\delta t}{\sum_{i=1}^{150} (E_{y\sim p_T}[y_{t_0+i}|y_{t_0}])^2 \delta t} \right\}^{1/2}.
\end{equation}
The integral error of the standard deviation is computed in the same method only by replacing the expectation by a standard deviation. The integral error of the expectation is only $e_\mu = 0.061$ and the standard deviation has an integral error of $e_{sd} = 0.016$.


\subsection{Mackey--Glass time series} \label{sec:MG}

\begin{figure}
  \centering
  \includegraphics[height=6cm]{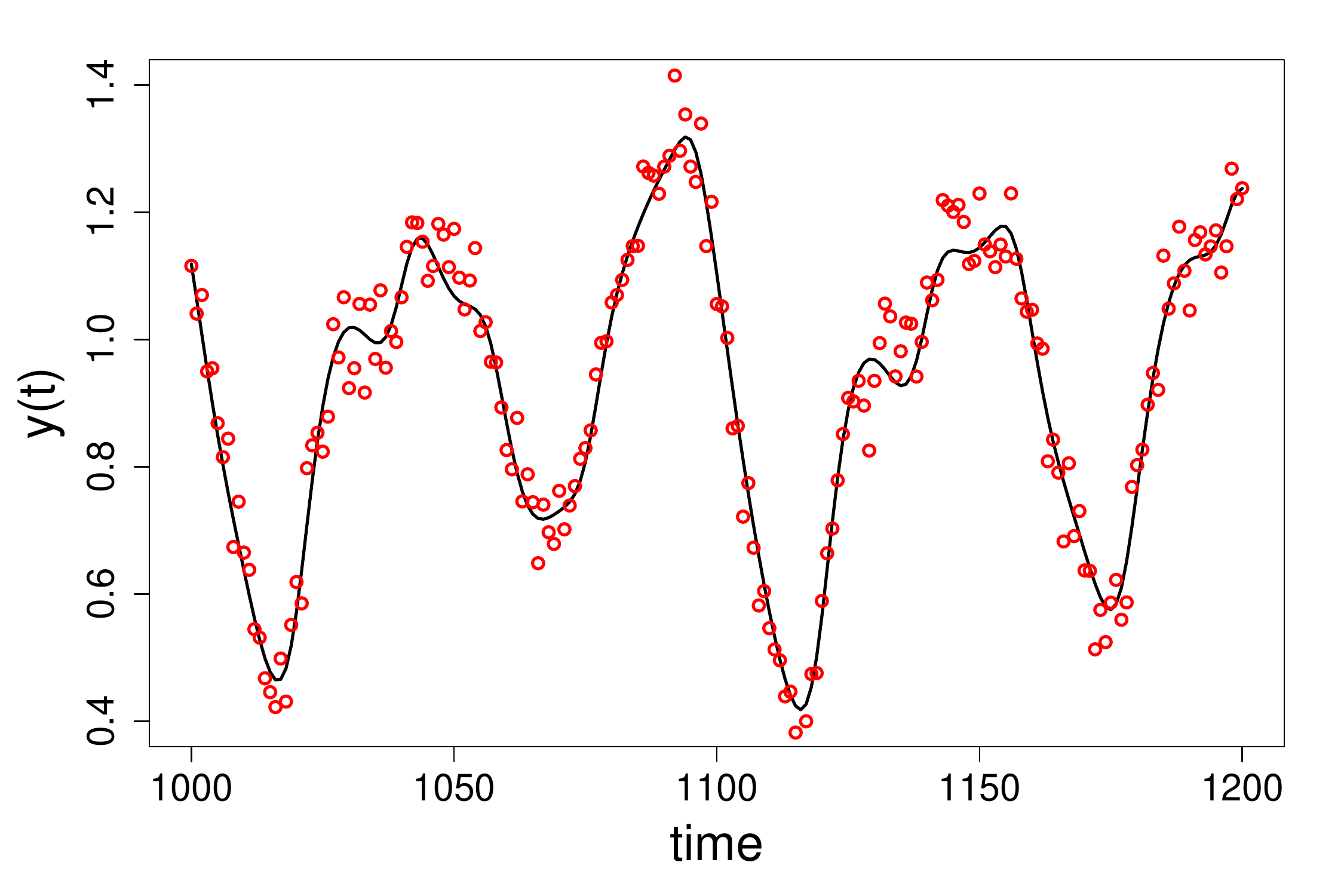}
  \caption{Noisy Mackey-Glass time series. The line is the ground truth, $y(t)$, and the circles denote the noisy data, $\yhat_t$. } \label{fig:MG_data}
\end{figure}

For the next test, DE-LSTM is used to simulate the Mackey-Glass time series \cite{Mackey77}. The Mackey-Glass equation is a nonlinear delay-time dynamical system, which has been extensively studied as a model chaotic system. The Mackey-Galss equation is
\begin{equation} \label{eqn:MG}
\frac{d y(t)}{dt} = \frac{\alpha y(t-\tau)}{1+y^\beta(t-\tau)} - \gamma y(t).
\end{equation}
We use the parameters adopted from \cite{Gers01}, $\alpha = 0.2$, $\beta = 10$, and $\gamma = 0.1$. For this set of parameters, the Mackey-Glass time series becomes chaotic for the time delay $\tau > 16.8$ \cite{Farmer82}. In this study, the time-delay parameter, $\tau = 17$, is used.

Equation (\ref{eqn:MG}) is numerically integrated by using a third-order Adams-Bashforth method with a time step size of 0.02, and a time series is generated by sampling $y(t)$ with a sampling interval, $\delta t = 1$. The time series data is corrupted by a white noise;
\[
\yhat_t = y(t) + \epsilon_t.
\]
The white noise is a zero-mean Gaussian random variable, $\epsilon_t \sim \mathcal{N}(0,\rho^2)$. The noise level is set to $\rho = 0.2 sd[y]$. The model training is performed similar to the previous experiments in section \ref{sec:OU_process}. A time series of the length $1.6 \times 10^5 \delta t$ is generated for the model training and another $2\times10^3 \delta t$ for the model validation. DE-LSTM is trained for $\delta y = 0.03 sd[\yhat]$. The grid space, $\delta y$, is chosen based on the noisy data. Assuming a uniform grid space, $\delta y$ should satisfy
\[
\delta y > \frac{\max(d\yhat)-\min(d\yhat)}{K},
\]
as the estimation interval, $\mathcal{I}_P = \alpha_{K+1} - \alpha_1$, should be larger than the support of $d\yhat$. In this example, $K=100$ is used, which makes $\delta y > 0.027 sd[\yhat]$.

\begin{figure}
  \centering
  \includegraphics[height=5cm]{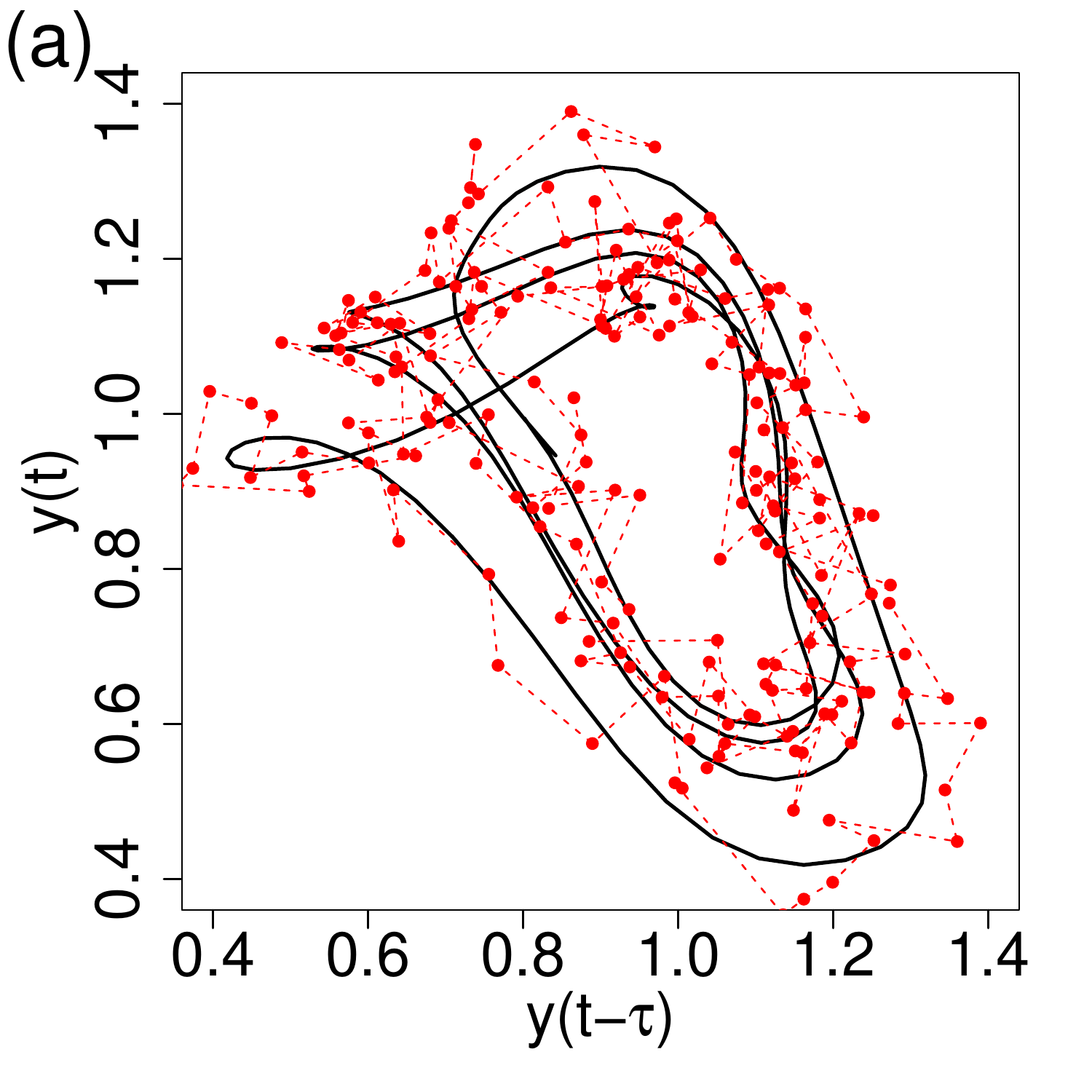}\hspace{1cm}
  \includegraphics[height=5cm]{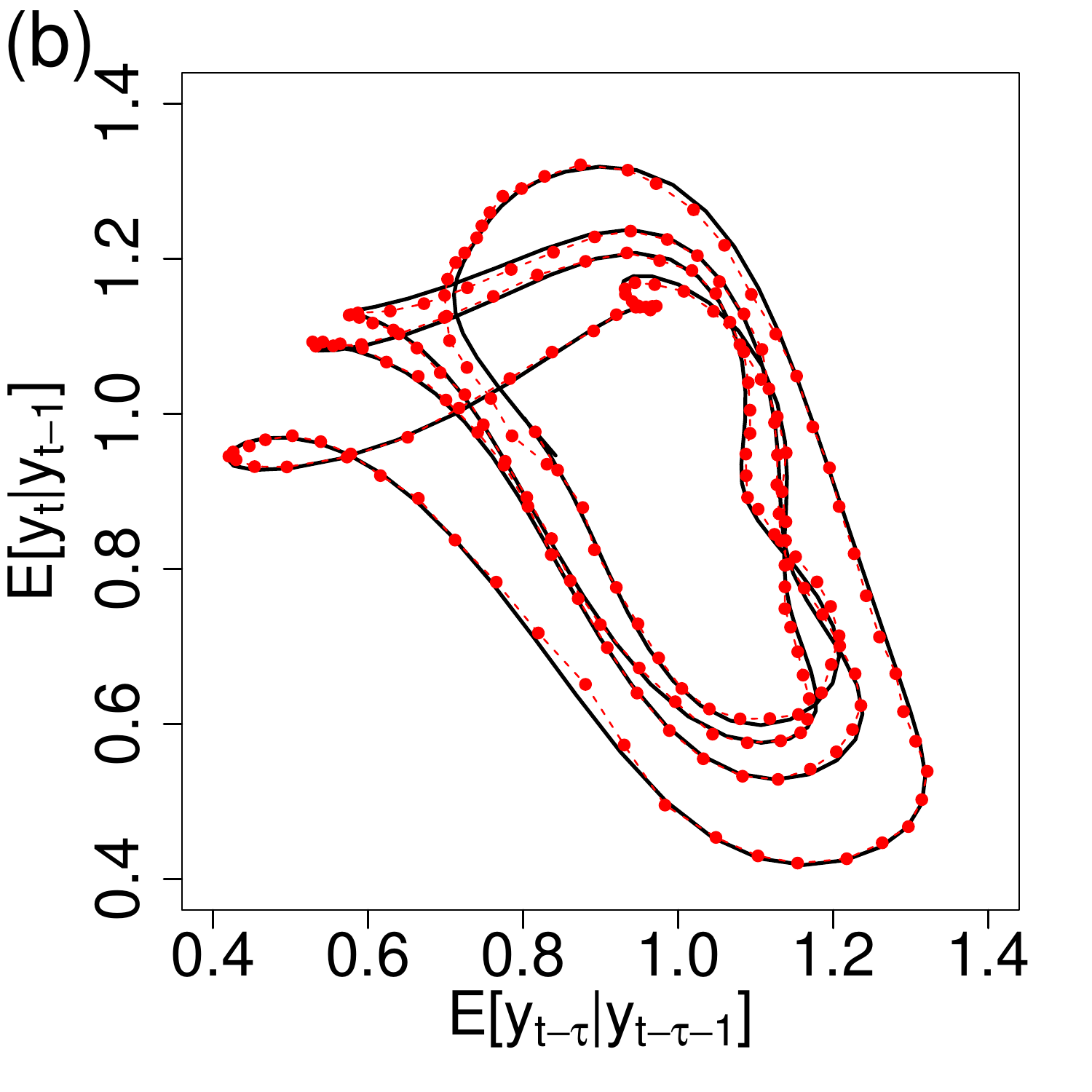}
  \caption{(a) Noisy observation in the delay-time phase space. The solid circles (${\color{red}\bullet}$) denote the noisy data and the ground truth is shown as a solid line ($\frac{~~~}{~~~}$). (b) The expectation of the next-step prediction by DE-LSTM is shown as the solid circles (${\color{red}\bullet}$). DE-LSTM is trained for $\delta y = 0.03 sd[\yhat]$ and $\lambda =  0.01$. } \label{fig:MG_phase}
\end{figure}

Figure \ref{fig:MG_phase} shows the expectation of the next-step prediction, $E[\yhat_{t+1}|\yhat_t]$, by DE-LSTM trained with $\lambda = 0.01$. Here, we slightly abuse the notation in the conditioning variables for simplicity. Instead of showing the dependence on the trajectories, only the last known data will appear in the notation, e.g., $E[\yhat_{t+1}|\yhat_t]$ for $E[\yhat_{t+1}|\widehat{\bm{Y}}_{0:t}]$. It is shown that when the noisy time series is presented to DE-LSTM (figure \ref{fig:MG_phase} a), DE-LSTM can effectively filter out the noise and reconstruct the original attractor of the Mackey-Glass system (figure \ref{fig:MG_phase} b). The root mean-square error of DE-LSTM with respect to the ground truth, $\langle (E[\yhat_{t+1}|\yhat_t] - y_{t+1} )^2 \rangle^{1/2}$, is only about 25\% of the noise level.

\begin{figure}
  \centering
  \includegraphics[height=5cm]{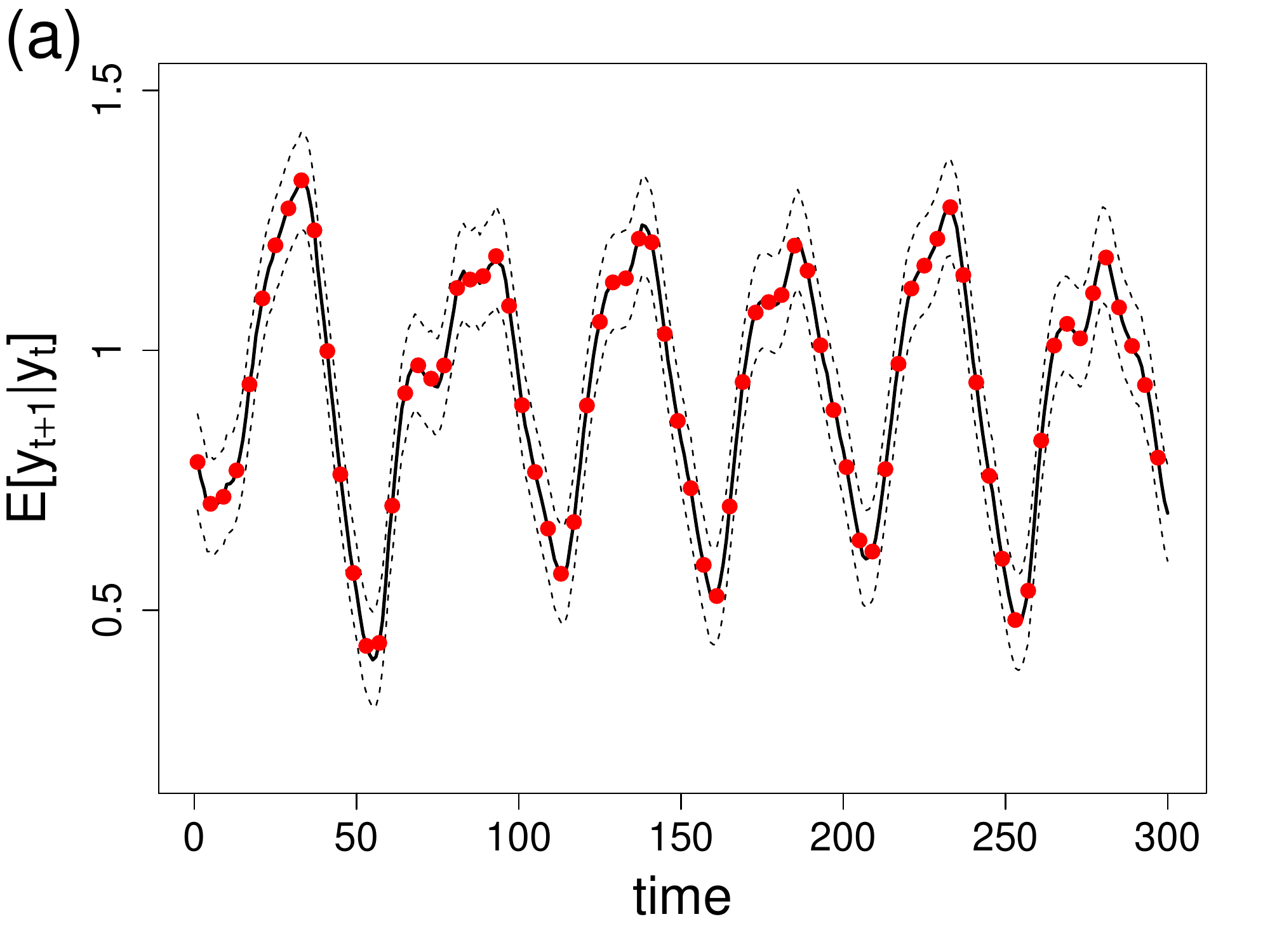}
  \includegraphics[height=5cm]{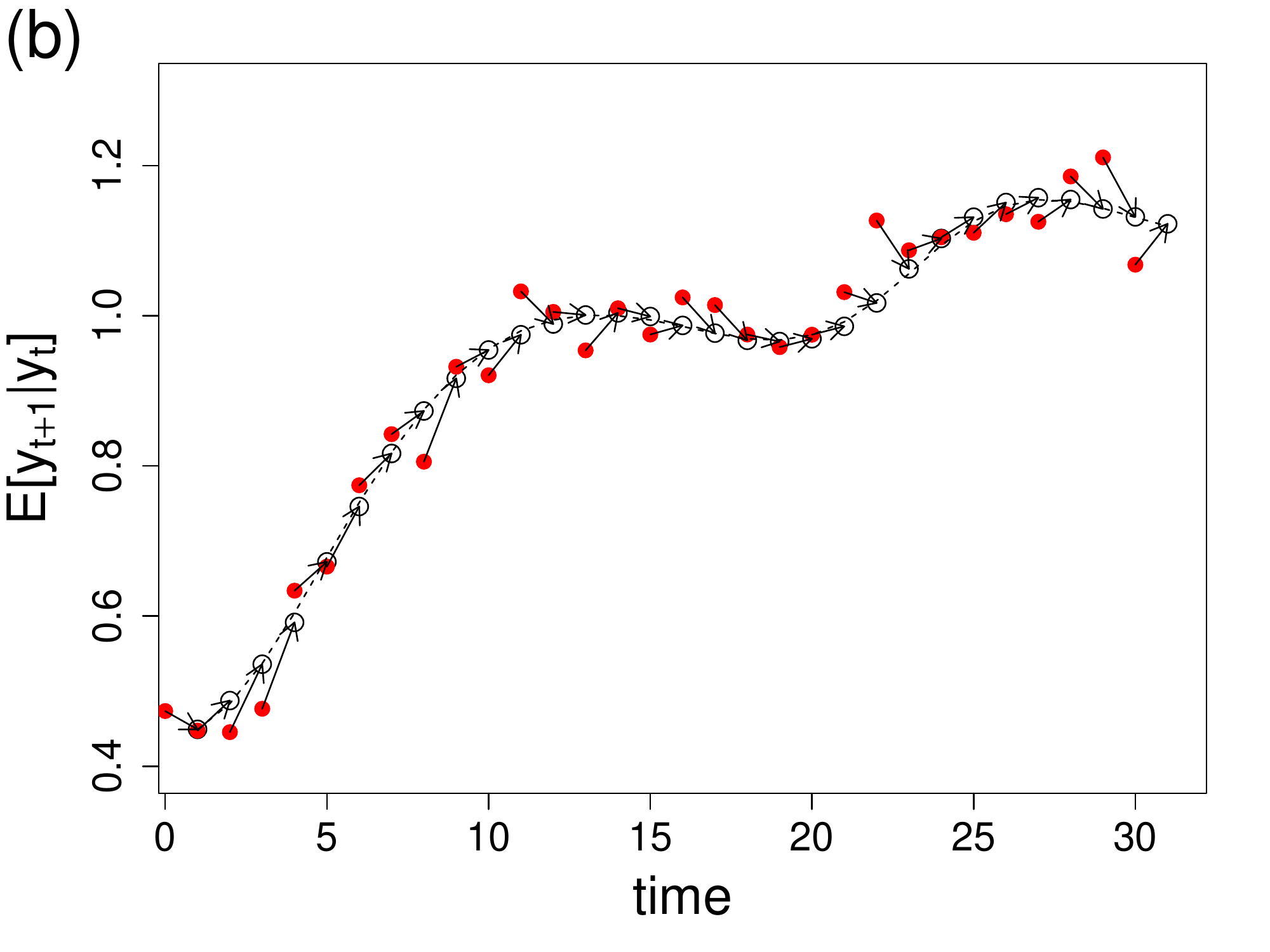}
  \caption{(a) Next-step prediction ($\frac{~~~}{~~~}$) and 95\% confidence interval ($\frac{~}{~}\frac{~}{~}\frac{~}{~}$) from DE-LSTM. The solid circles denote the ground truth, $y(t)$. (b) DE-LSTM prediction ($\circ$) from the noisy data ({\color{red}$\bullet$}). The dashed line is the ground truth, $y(t)$. DE-LSTM is trained for $\delta y = 0.03 sd[\yhat]$ and $\lambda =  0.01$. } \label{fig:MG_pred}
\end{figure}

Figure \ref{fig:MG_pred} (a) shows the next-step prediction of DE-LSTM. Figure \ref{fig:MG_pred} (b) illustrates the actual prediction process. At every $t$, a noisy data, $\yhat_t$, is shown to DE-LSTM. Then, DE-LSTM updates its internal state and predict the probability distribution of $d\yhat$, $p(d \yhat_t|\yhat_t)$. The expectation at $t+1$ is computed as $E[\yhat_{t+1}|\yhat_t] = \yhat_t + E[d\yhat_t | \yhat_t]$. Hence, to make a correct prediction, DE-LSTM should be able to know how far away $\yhat_t$ is from the true state, $y_t$. 

Here, we define a normalized root mean-square error as
\begin{align} \label{eqn:mg_nrmse}
e_\mu &= \frac{\langle ( E[\yhat_{t+1}|\yhat_t] - y_{t+1} )^2 \rangle^{1/2}}{\langle (y_{t+1} - \yhat_t)^2 \rangle^{1/2}} =  \frac{\langle ( E[\yhat_{t+1}|\yhat_t] - y_{t+1} )^2 \rangle^{1/2}}{\left( \langle dy^2_t \rangle + \langle \epsilon_t^2 \rangle \right)^{1/2}},\\
e_{sd} &= \frac{\langle Var[d\yhat_t|\yhat_t] \rangle^{1/2}}{\rho} - 1,
\end{align}
in which $dy_t = y(t+\delta t) - y(t)$. The normalized root mean-square errors are listed in table \ref{tbl:MG_error}. Similar to the Ornstein-Uhlenbeck process, $e_\mu$ decreases at first for a smaller value of $\lambda$ and, then, starts to increase for $\lambda \ge 0.1$. It is shown that $e_{sd}$ become larger as $\lambda$ increases. Because a larger value of $\lambda$ oversmooths the probability distribution, $e_{sd}$ becomes an increasing function of $\lambda$. 

\begin{table}
\center{
\caption{Normalized root mean-square errors of the next-step prediction for DE-LSTM, LSTM, autoregressive integrated moving average (ARIMA), and Kalman filter (KF). DE-LSTM is trained with $\delta y = 0.03sd[\yhat]$.} \label{tbl:MG_error}
\begin{tabular}{c|cccc|c|c|c}
\hline \hline 
\multirow{2}{*}{ } & \multicolumn{4}{c|}{$\lambda$} & \multirow{2}{*}{LSTM}&\multirow{2}{*}{ARIMA}&\multirow{2}{*}{KF} \\
           &       0 & 0.001 &   0.01 &     0.1 &           &           &          \\
\hline
$e_\mu$& 0.186 & 0.179 & 0.170 & 0.196 & 0.177 & 1.403 & 1.441 \\
$e_{sd}$& 0.023 & 0.028 & 0.043 & 0.151 &     -    & 0.531 & 0.517 \\
\hline\hline
\end{tabular}
}
\end{table}

For a comparison, a standard (regression) LSTM and two most widely used time series prediction models, auto-regressive integrated moving average (ARIMA) and Kalman filter (KF), are also trained against the same data. 
{
The ``forecast'' package for the {\bf R} system for statistical computing is used to build the ARIMA model \cite{Hyndman08}. The model parameters are chosen by using the Akaike's Information Criterion \cite{ Box08,Sakamoto87}. The ``dse'' package is used for the Kalman filter \cite{Gilbert06}, where the model parameters are estimated by the maximum likelihood method \cite{Bishop06}.}
The results are shown in table \ref{tbl:MG_error}.As expected, the prediction from the regression LSTM is as good as that from DE-LSTM. However, it should be noted that the regression LSTM only provides a deterministic prediction, while DE-LSTM provides richer information about the probability distribution of the prediction. It is shown that $e_\mu$ of ARIMA and KF are much larger than DE-LSTM. 
Since the Mackey-Glass equation is a delay-time nonlinear dynamical system, it is not surprising that those linear models are not able to make good predictions. Moreover, the delay-time parameter is $\tau = 17 \delta t$, indicating that the model should be able to conserve the state of the system at 17 time-step ago and to use the information at a correct timing. Such a long time dependence is very difficult to incorporate without a prior knowledge in the conventional time series models. But, the experimental results suggest that DE-LSTM can comprehend such a long time dependence without any prior information.

\begin{table}
\center{
\caption{Kullback-Leibler divergence in terms of $\lambda$.} \label{tbl:MG_KL}
\begin{tabular}{c|cccc}
\hline \hline
$\lambda$ & 0 & $10^{-3}$ & $10^{-2}$ & $10^{-1}$ \\
\hline
$D_{KL}~(\times 10^3)$ & $1.13$ & $0.75$ & $0.68$ & $1.24$ \\
\hline \hline
\end{tabular}
}
\end{table}

Table \ref{tbl:MG_KL} shows the Kullback--Leibler divergence in terms of $\lambda$. In table \ref{tbl:MG_error}, $e_{sd}$ is smallest at $\lambda = 0$ and is shown to be an increasing function of $\lambda$. However, when the predicted probability distribution is directly compared with the ground truth by using the Kullback-Leibler divergence, it is shown that $D_{KL}$ at $\lambda = 0$ is larger than those at $\lambda = 0.001$ and 0.01. Without the smoothness constraint, DE-LSTM still matches the first and second moments of the probability distribution, but the computed probability distribution is very bump, which makes $D_{KL}$ larger.

\begin{figure}
  \centering
  \includegraphics[width=0.95\textwidth]{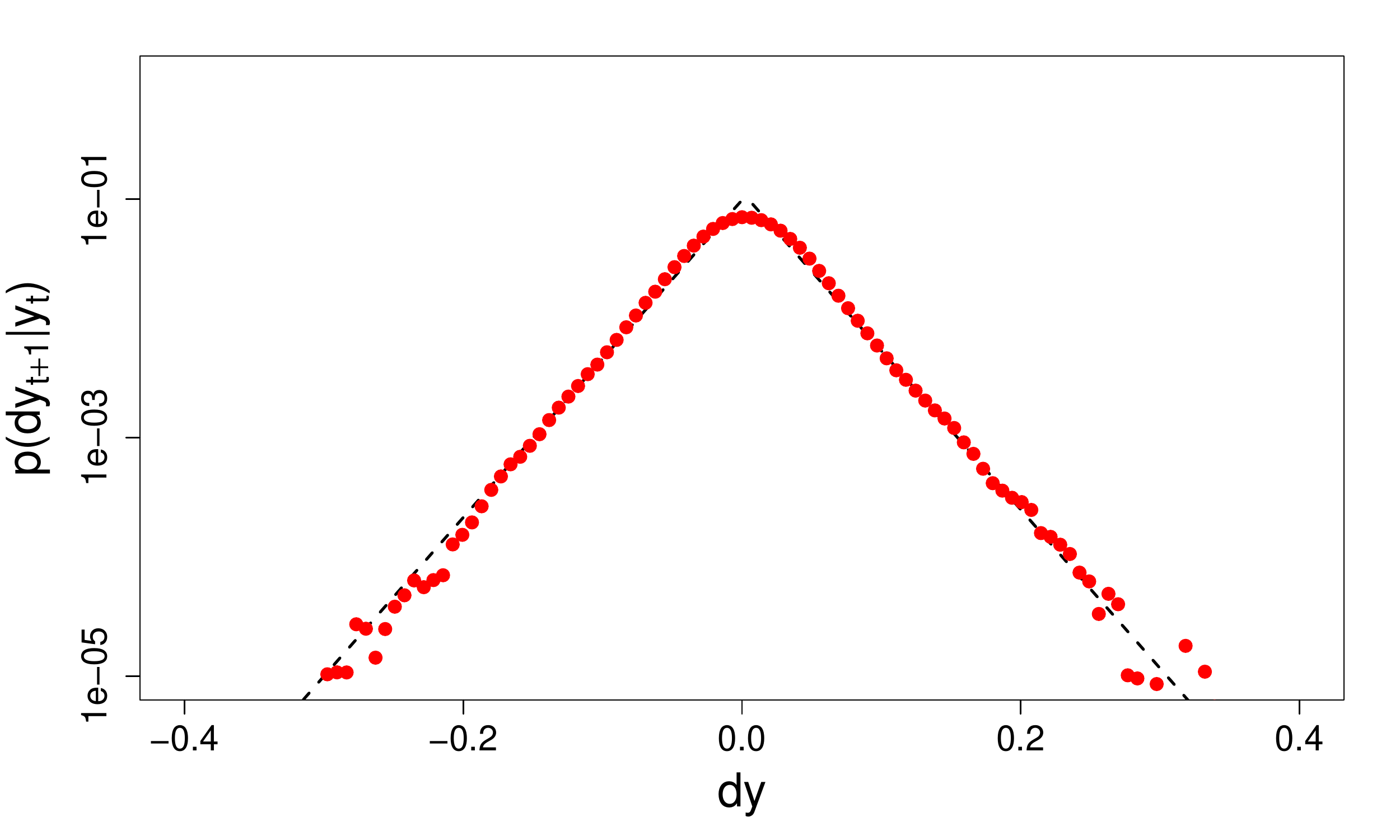}
  \caption{Next-step prediction of the probability distribution, $p(\yhat_{t+1}|\yhat_t)$, for the Laplace noise; the circles (${\color{red}\bullet}$) show DE-LSTM and the dashed line is the ground truth.}\label{fig:MG_Laplace}
\end{figure}

In figure \ref{fig:MG_Laplace}, DE-LSTM is tested for a Laplace noise. The probability density function of $\epsilon_t$ is given as
\begin{equation}
p(\epsilon) = \frac{1}{2b}\exp\left(-\frac{|\epsilon|}{b}\right),
\end{equation}
in which $b=0.2sd[y]/\sqrt{2}$. The simulation parameters are not changed for the Laplace noise; $\delta y = 0.03 sd[\yhat]$ and $\lambda = 0.01$. It is shown that DE-LSTM well captures the Laplace distribution without assuming a distributional property of the noise.

\begin{table}
\center{
\caption{Kullback-Leibler divergence in terms of the number of the LSTM units ($N_c$). The penalty parameter is fixed at $\lambda = 10^{-2}$.} \label{tbl:MG_KL_Nc}
\begin{tabular}{c|cccccc}
\hline \hline
$N_c$ &16 & 32 & 64 & 128 & 256 & 512 \\
\hline
$D_{KL}~(\times 10^3)$ & $5.28$ & $1.10$ & $0.70$ & $0.68$ & $0.67$ & $0.87$\\
\hline \hline
\end{tabular}
}
\end{table}

To investigate the effects of the number of LSTM units, $N_c$, DE-LSTM is trained for a range of $N_c$, while all other parameters are fixed, e.g., $\delta y = 0.03$ and $\lambda = 0.01$. Table \ref{tbl:MG_KL_Nc} shows the Kullback-Leibler divergence as a function of $N_c$. In general, $D_{KL}$ decreases as more LSTM units are used. There is a noticeable drop in $D_{KL}$ when $N_c$ is changed from 16 to 64. But, for $Nc = 64 \sim 256$, $D_{KL}$ becomes essentially flat. When $N_c$ is increased further, to 512 LSTM units, $D_{KL}$ starts to grow slowly. This results suggest that, for the Mackey--Glass time series, $Nc = 64$ is sufficient to represent the dynamics and adding more LSTM units beyond the threshold does not necessarily improve the accuracy of DE-LSTM. It is also observed that, when too many LSTM units are used ($N_c = 512$), the accuracy is getting worse. The representation capability of an artificial neural network becomes more powerful as more LSTM units are added. However, as the number of parameters increases, it becomes increasingly more difficult to train a neural network given a fixed size data. In this example, the number of parameters of DE-LSTM changes from 48,293 at $N_c = 64$ to 2,679,397 at $N_c = 512$. 

\begin{table}
\center{
\caption{{
Wall-clock computation time for the next-step prediction in seconds. $N_c$ denotes the number of the LSTM units.}} \label{tbl:pred_wallclock}
\begin{tabular}{c|ccccc}
\hline \hline
$N_c$ &32 & 64 & 128 & 256 & 512 \\
\hline
3-CPU ($\times 10^4$) & $1.8$ & $2.6$ & $3.7$ & $8.7$ & $32.0$\\
1-GPU ($\times 10^4$) & $4.8$ & $4.5$ & $4.7$ & $5.0$ & $4.9$\\
\hline \hline
\end{tabular}
}
\end{table}

{
In this study, the computations are performed by using \emph{Torch} \cite{torch}. Table \ref{tbl:pred_wallclock} shows the wall-clock computation times of the next-step prediction either on CPUs (Intel Xeon E5-2620 2.10GHz) or a GPU (Nvidia Tesla K80). The wall-clock time is measured by averaging over the LSTM simulation for 100 time steps. When the CPUs are used for the computation, the maximum number of OpenMP threads is set to three. For a small LSTM network, $N_c < 256$, using CPUs with OpenMP is faster than GPU computation. But, as $N_c$ increases, the GPU outperforms the CPU computation. It is shown that, for the range of $N_c$ used in this study, the wall-clock computation time on the GPU is essentially unchanged from $5.0\times10^{-4}$ seconds. Computing one-step of an LSTM network consists of many small-size matrix-vector or matrix-matrix operations. Hence, when the size of the LSTM network is small, the overhead overwhelms the computation time, which explains why the wall-clock time of the GPU computation does not increase with the number of the LSTM units.
}

\begin{figure}
  \centering
  \includegraphics[width=0.95\textwidth]{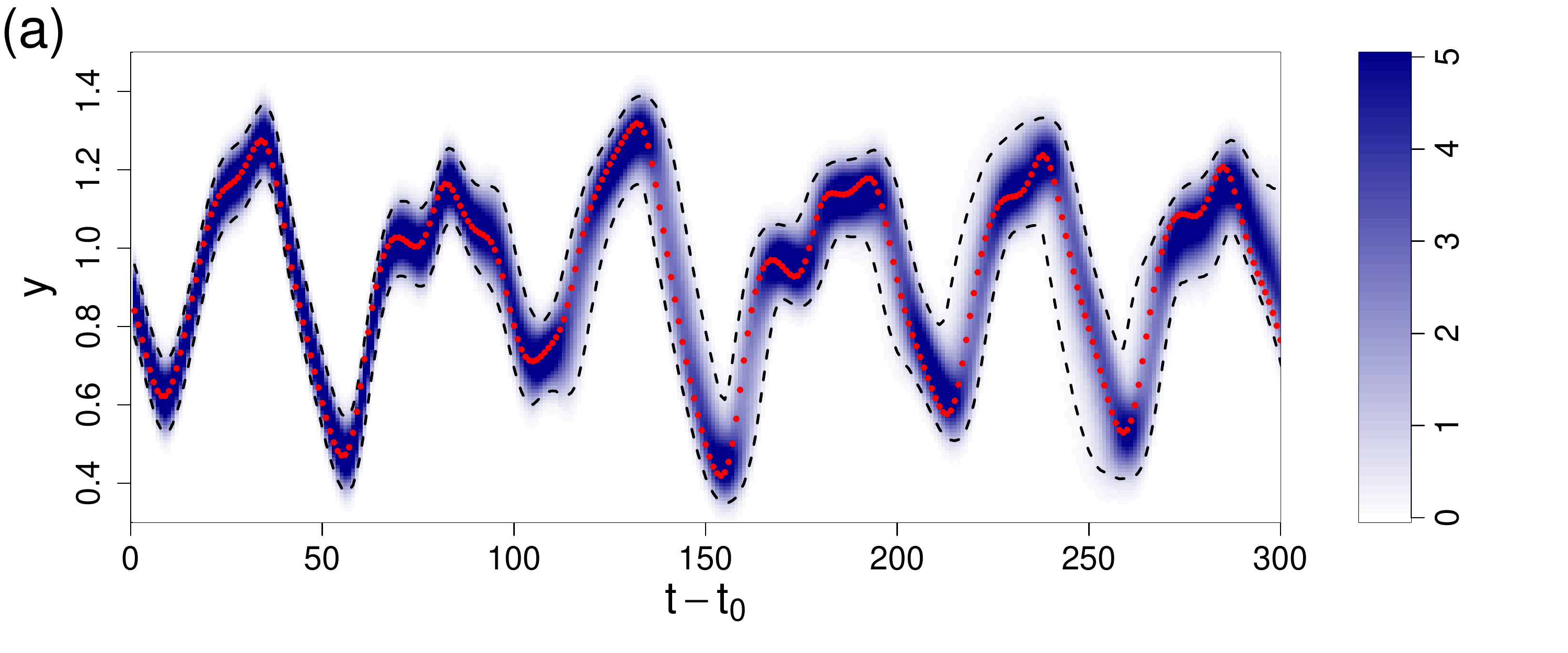}\\
  \includegraphics[width=0.95\textwidth]{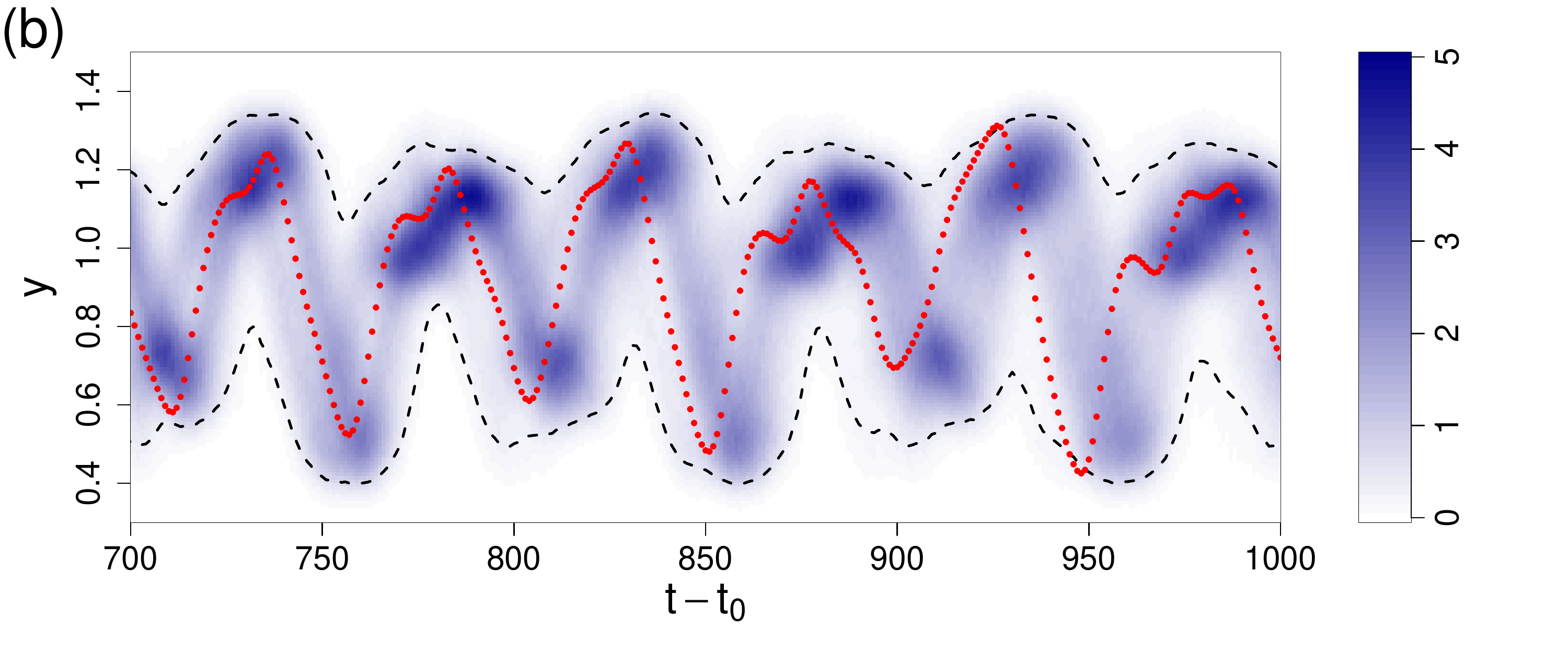}
  \caption{1,000-step forecast of Mackey-Glass time series by DE-LSTM. The dashed lines are 95\% confidence interval and the solid circles denote the ground truth, $y(t)$. The color contours denote the probability density function, $p(\yhat_{t_0+n}|\yhat_{t_0})$.}\label{fig:MG_multi_step_dist}
\end{figure}

A multiple-step forecast of the Mackey-Glass time series is shown in figure \ref{fig:MG_multi_step_dist}. The noisy observation, $\yhat_t$, is supplied to DE-LSTM for the first 100 steps ($t = - 99 \sim 0$) for an initial spin-up of the internal states, and the time evolution of the probability distribution, $p(\yhat_{t_0+n}|\yhat_{t_0})$, is computed for the next 1,000 time steps by using $2 \times 10^4$ Monte Carlo samples. The probability density function is estimated by a kernel density estimation. A Gaussian kernel, of which bin size is $\delta y_{kde} = 0.04sd[\yhat]$ and kernel width $\sigma_{kde} = 0.06sd[\yhat]$, is used for the density estimation. The Mackey--Glass time series in this experiment has a characteristic period of $T_c \simeq 50$ \cite{Gers01}. The forecast horizon corresponds to about 20 $T_c$. Since the Mackey--Glass time series is chaotic, a deterministic long-term forecast is impossible when the initial condition is given as a random variable, $\yhat_t$. However, in the probabilistic forecast by DE-LSTM, it is shown that the 95\% confidence interval (95-CI) encompasses the ground truth even when the forecast horizon is $t > 700$ (figure \ref{fig:MG_multi_step_dist} b). To quantify the accuracy of the multiple-step forecast, we define an empirical coverage probability,
\begin{equation} \label{eqn:coverage}
\mathcal{B} = \frac{1}{N} \sum_{t=1}^{N} \chi_{CI}(\yhat_{t_0+t}).
\end{equation}
Here, $N$ is a forecast horizon, e.g., $N = 1000$, and $\chi_{CI}(\yhat_t)$ is an indicator function, which is one if $\yhat_t$ is within 95-CI and zero otherwise. In this example, it is found that the coverage probability is $\mathcal{B} = 0.944$, suggesting the forecast of the future probability distribution is reliable.


%

\begin{figure}
  \centering
  \includegraphics[height=4.5cm]{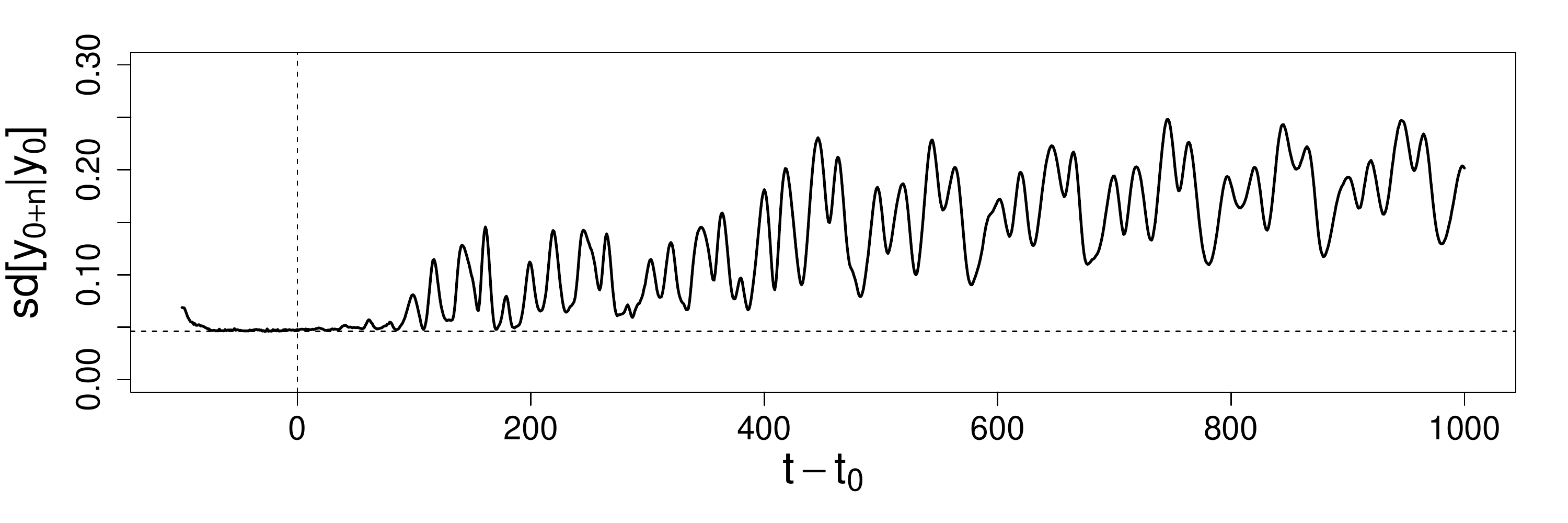}
  \caption{Temporal evolution of the standard deviation of $\yhat_{t+n}|\yhat_t$. The horizontal dashed line denotes the noise level, $\rho$.}\label{fig:MG_multi_step_std}
\end{figure}

The multiple-step forecast corresponds to propagating uncertainty in time. In a conventional linear time series model, the forecast uncertainty is usually a non-decreasing function of 
{
the forecast horizon \cite{Box08}}. However, in DE-LSTM, it is shown that the forecast uncertainty, e.g., 95-CI, is no longer a monotonic function of time. Even at a long forecast horizon (figure \ref{fig:MG_multi_step_dist} b), 95-CI dynamically adjusts with the period of the Mackey--Glass time series. Figure \ref{fig:MG_multi_step_std} shows the temporal evolution of the standard deviation of $p(\yhat_{t_0+n}|\yhat_{t_0})$. It is shown that, after a short transient state ($-100<t<-80$), the estimated standard deviation, $sd[\yhat_{t+1}|\yhat_t]$, becomes close to the noise level, $\rho = 0.2 sd[y]$. When the multiple-step prediction is started at $t=0$, the standard deviation, $sd[\yhat_{t_0+n}|\yhat_{t_0}]$, still remains close to $\rho$ until $t \simeq 80$, then starts to grow for larger $t$. It is interesting to observe that, in DE-LSTM, $sd[\yhat_{t_0+n}|\yhat_{t_0}]$ exhibits a nonlinear behavior in time. The forecast standard deviation, $sd[\yhat_{t_0+n}|\yhat_{t_0}]$, exhibits very large fluctuations, indicating that the prediction uncertainty may increase or decrease following the dynamics of the system. For $t > 600$, $sd[\yhat_{t_0+n}|\yhat_{t_0}]$ still oscillates, but no longer grows in time.

\begin{table}
\center{
\caption{{Wall-clock computation time of one time step of the Monte Carlo simulation in seconds on GPU. $N_s$ denotes the number of the Monte Carlo samples.}} \label{tbl:MC_wallclock}
\begin{tabular}{c|ccccc}
\hline \hline
$N_s$ ($\times 10^{-3}$) &1 & 2 & 4 & 8 & 16 \\
\hline
LSTM ($\times 10^4$) & 6.5 & 6.7 & 7.5 & 8.2 & 8.7\\
Sampling & 0.03 & 0.06 & 0.12 & 0.24 & 0.48\\
\hline \hline
\end{tabular}
}
\end{table}

{
Table \ref{tbl:MC_wallclock} shows the computation time of the Monte Carlo simulation on the GPU (Nvidia Tesla K80). The computation time is divided into two categories; the computation of the LSTM network (LSTM) and the sampling from $\bm{P}^{(i)}_t$ for $i=1,\cdots,N_s$. It is shown that the increase in the wall-clock time of the LSTM computation is very mild. When the number of the Monte Carlo samples is increased from 1,000 to 16,000, the wall-clock time changes only from $6.5\times10^{-4}$ to $8.7\times10^{-4}$ seconds per one time step. The wall-clock time is computed by averaging over 1,000 time steps. 
In this example, most of the computation time is spent in drawing samples from the predicted probability distribution. For $N_s = 16,000$, the wall-clock time of the sampling procedure is about 550 times larger than the LSTM computation time. Because the sampling is performed sequentially in this study, it is shown that the wall-clock time of the sampling procedure increases linearly with $N_s$. If a parallel computing technique is employed in the sampling procedure, it is expected that the wall-clock time of the Monte Carlo simulation can be drastically reduced. 
}

\subsection{Forced Van der Pol oscillator} \label{sec:VDP}

\begin{figure}
  \centering
  \includegraphics[height=4.5cm]{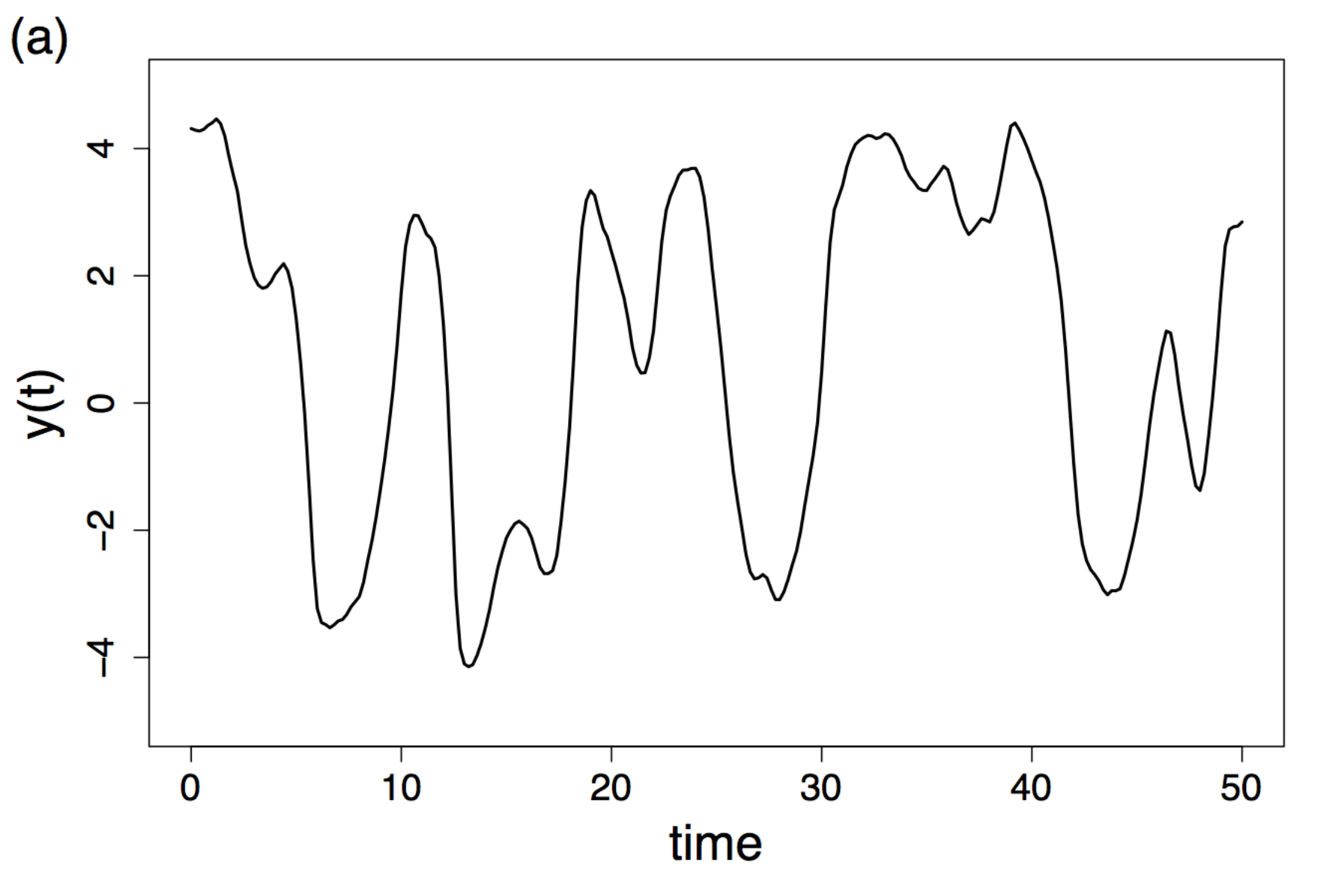}
  \includegraphics[height=4.5cm]{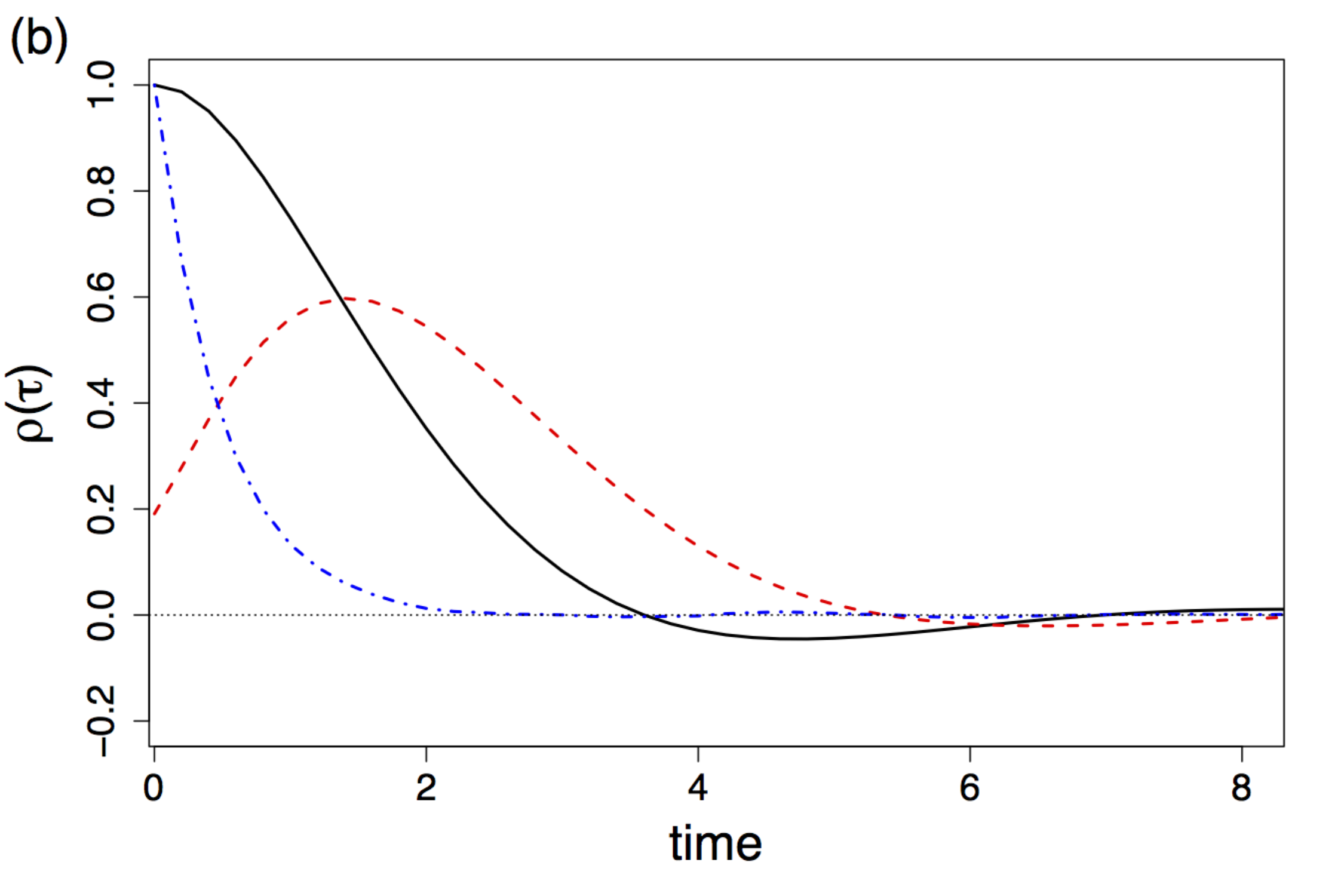}
  \caption{(a) Sample trajectory of the forced Van der Pol oscillator and (b) auto-correlation functions; $\frac{~~~~~}{~~~~~}$, $\rho(\tau;y,y)$; {\color{blue}$\frac{\,~}{\,~}\,\frac{\,}{\,}\,\frac{\,~}{\,~}$}, $\rho(\tau;u,u)$; {\color{red}$\frac{~\,}{~}\,\frac{~\,}{~}\,\frac{~\,}{~}$}, $\rho(\tau;u,y)$.}\label{fig:VDP_simul} 
\end{figure}

The next example is a forced Van der Pol oscillator (VDP), which is given by the following equations,
\begin{align}
\frac{dy_1}{dt} &= y_2, \label{eqn:VDP_1}\\
\frac{dy_2}{dt} &= \alpha (1-y_1^2)y_2 - y_1 + u(t) \label{eqn:VDP_2}
\end{align} 
The exogenous forcing, $u(t)$, is given by an Ornstein-Uhlenbeck process as
\begin{equation} \label{eqn:VDP_OU}
du = -\theta u dt + \xi dW.
\end{equation}
The parameters used in this simulation are, $\alpha = 0.5$, $\theta = 2$, and $\xi = 5\sqrt{2\theta}$. After solving equations (\ref{eqn:VDP_1}--\ref{eqn:VDP_2}), only $y_1(t)$ is provided to DE-LSTM as a target variable, i.e., $y_t = y_1(t)$ and $\bm{x}_t = (y_1(t),u(t))$. Figure \ref{fig:VDP_simul} (a) shows a sample trajectory of $y_1(t)$. In figure \ref{fig:VDP_simul} (b), the auto-correlation functions are displayed. The auto-correlation functions are defined as
\begin{equation}
\rho(\tau;a,b) = \frac{\langle a(t)b(t+\tau) \rangle}{\langle a^2(t)\rangle^{1/2}\langle b^2(t)\rangle^{1/2}}.
\end{equation}
The auto-correlation functions indicate that the relaxation timescales of $y(t)$ and $u(t)$ are different from each other. And, the cross auto-correlation, $\rho(\tau;u,y)$, shows that VDP is also a delay-time dynamical system, in which the effects of $u(t)$ on $y(t)$ becomes maximum after some time delay, $\tau \simeq 1.8$.

The Van der Pol oscillator is numerically solved by using a third-order Adams-Bashforth method with a time step size of 0.001, and a time series is generated by sampling $y(t)$ with a sampling interval, $\delta t = 0.2$. The time series data is corrupted by a white noise;
\[
\yhat_t = y_t + \epsilon_t.
\]
The white noise is a zero-mean Gaussian random variable, $\epsilon_t \sim \mathcal{N}(0,\rho^2)$ with the standard deviation $\rho = 0.2 sd[y]$. A time series of the length $1.6 \times 10^5 \delta t$ is generated for the model training and another $2\times10^3 \delta t$ for the model validation. DE-LSTM is trained for $\delta y = 0.03 sd[\yhat]$ and $\lambda = 0.1$.

\begin{figure}
  \centering
  \includegraphics[height=5cm]{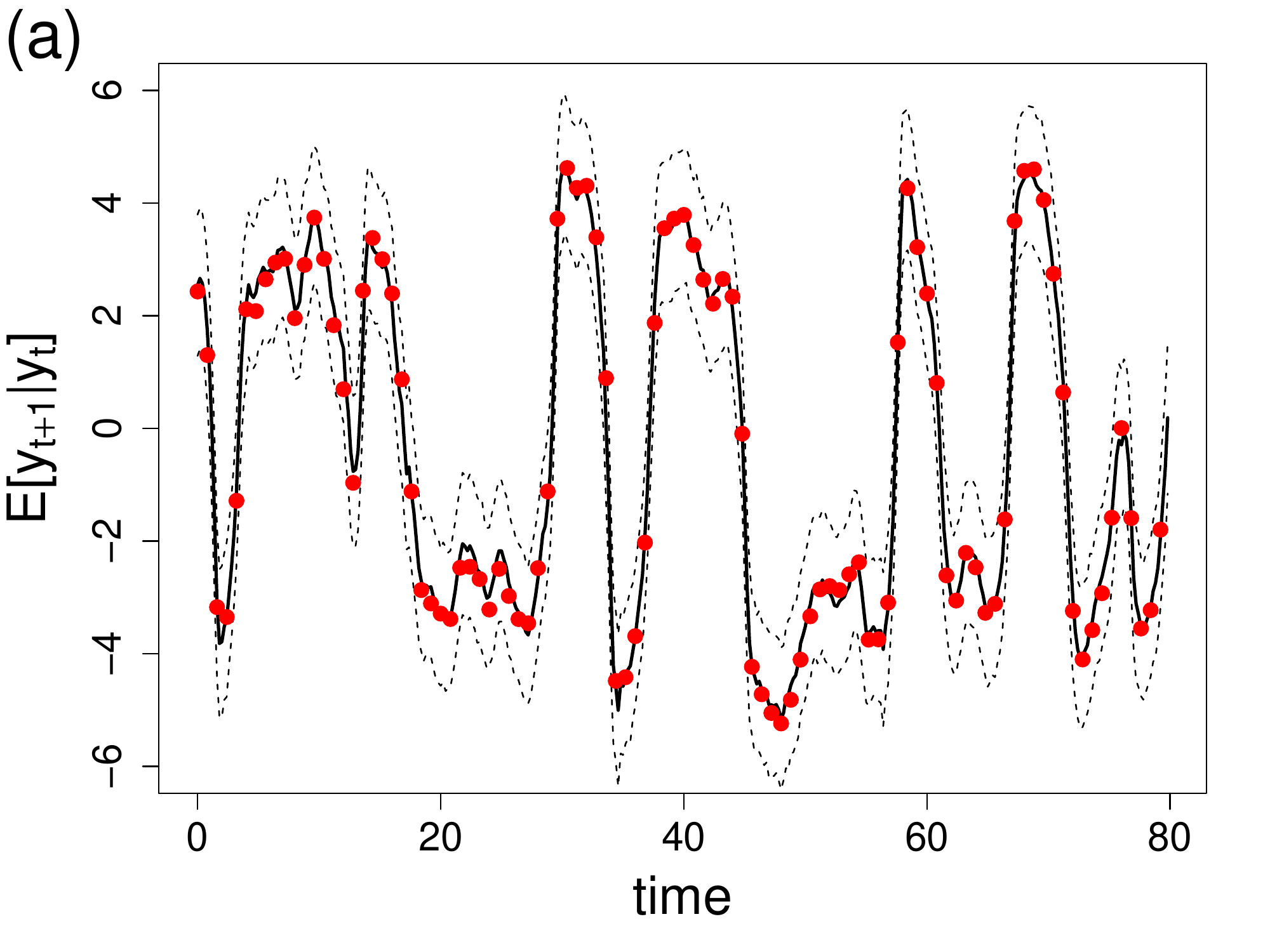}
  \includegraphics[height=5cm]{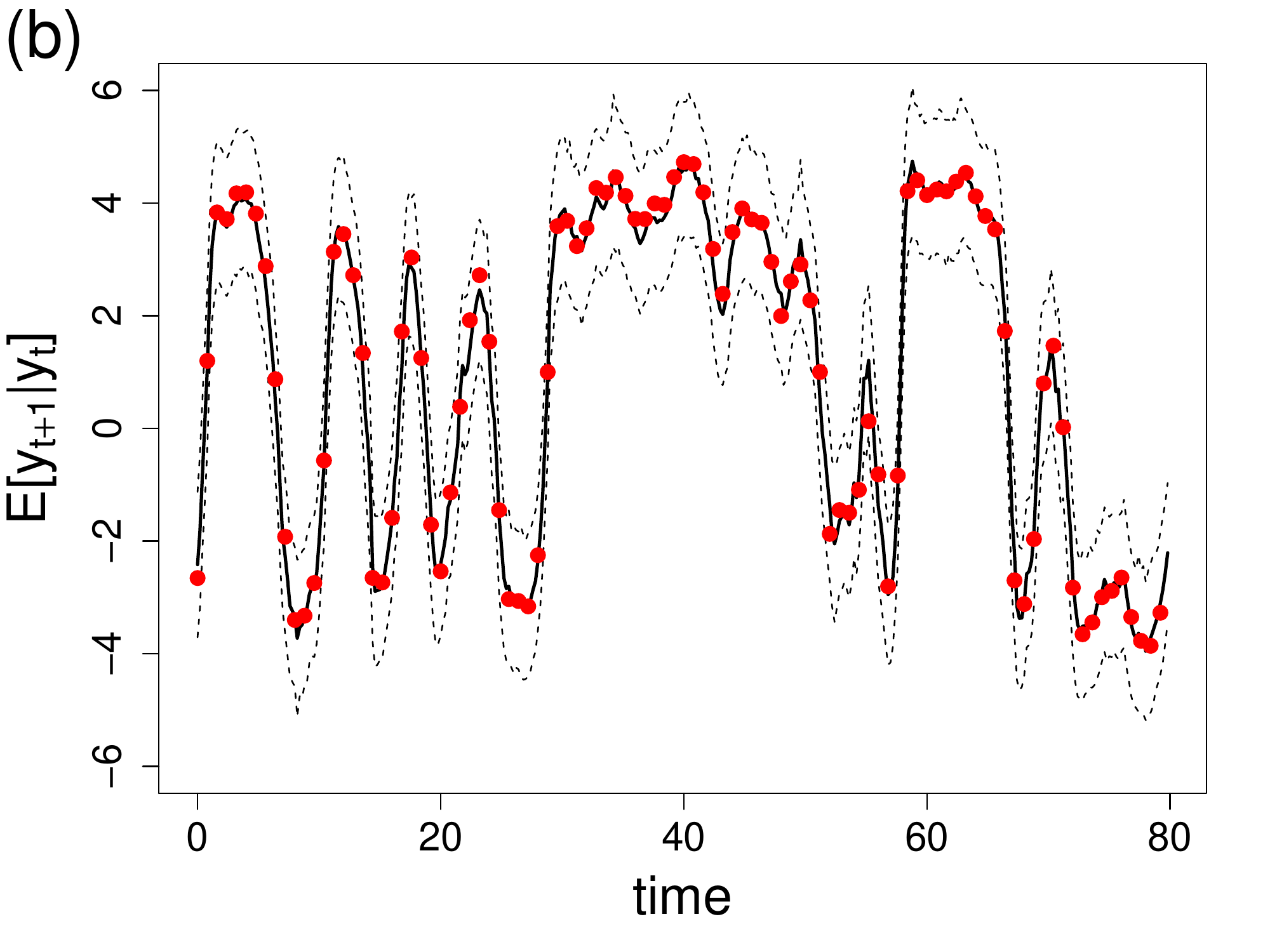}
  \caption{Next-step predictions ($\frac{~~~}{~~~}$) and 95\% confidence intervals ($\frac{~}{~}\frac{~}{~}\frac{~}{~}$) from DE-LSTM for two different exogenous forcing; (a) $\theta = 2$ and (b) $\theta=1$. The solid circles denote the ground truth, $y(t)$. DE-LSTM is trained by using $\theta = 2$.  } \label{fig:VDP_pred}
\end{figure}

Figure \ref{fig:VDP_pred} (a) shows the next-step prediction by DE-LSTM. Similar to the Mackey-Glass equations, DE-LSTM makes a very good prediction of the ground truth, $y(t)$. The root mean-square error of the expectation is less than the noise level, $\langle (E[\yhat_{t+1}|\yhat_t] - y_{t+1})^2 \rangle^{1/2} \simeq 0.44 sd[\epsilon_t]$. 

\begin{table}
\center{
\caption{Normalized root mean-square errors of the next-step prediction for DE-LSTM, autoregressive integrated moving average (ARIMA), and Kalman filter (KF).} \label{tbl:VDP_error}
\begin{tabular}{c|c|ccc}
\hline \hline 
&$\theta = 2$ & \multicolumn{3}{c}{$\theta = 1$} \\
\hline
&DE-LSTM & DE-LSTM & ARIMA & KF\\
\hline
$e_\mu$ & 0.33 & 0.30 & 1.24 & 1.18\\
$e_{sd}$ & 0.20 & 0.04 & 0.42 & 0.42\\
\hline\hline
\end{tabular}
}
\end{table}

One of the speculations on deep learning is that, due to the large number of parameters, a deep neural network can simply memorize the training sequence, instead of learning the representation \cite{Zhang_ICLR_17}. When a deep neural network is used to model a discrete data with a finite number of possible states, the hypothesis is plausible. But, in modeling a continuous time series, the space spanned by the time series data is too big for a deep neural network to memorize. To test the representation capability of DE-LSTM, a new forced VDP data set is generated with a different exogenous forcing. For the new exogenous forcing, $u(t)$, the same Ornstein-Uhlenbeck process (\ref{eqn:VDP_OU}) is used, but the timescale parameter is changed from $\theta = 2$ to $\theta =1$ and the variance of the Weiner process, $\xi=5\sqrt{2\theta}$, is also changed accordingly. Then, DE-LSTM trained using the original data set ($\theta = 2$) is used to make a prediction of the new data ($\theta=1$). The next-step prediction for this new VDP is shown in figure \ref{fig:VDP_pred} (b). Note that the same DE-LSTM is used for the predictions in figure \ref{fig:VDP_pred} (a) and (b). Although DE-LSTM is tested against the VDP dynamical system with a different timescale, it stills makes a very good prediction. Table \ref{tbl:VDP_error} shows $e_\mu$ and $e_{sd}$ of DE-LSTM, ARIMA, and KF. It is shown that NRMSEs of DE-LSTM do not change much from $\theta = 1$ to $\theta =2$. As expected, NRMSEs of ARIMA and KF are much larger than DE-LSTM due to the nonlinearity of VDP. This result suggests that DE-LSTM can learn a representation for the nonlinear dynamical system, rather than just memorizing the input-output sequence.

\begin{figure}
  \centering
  \includegraphics[width=0.95\textwidth]{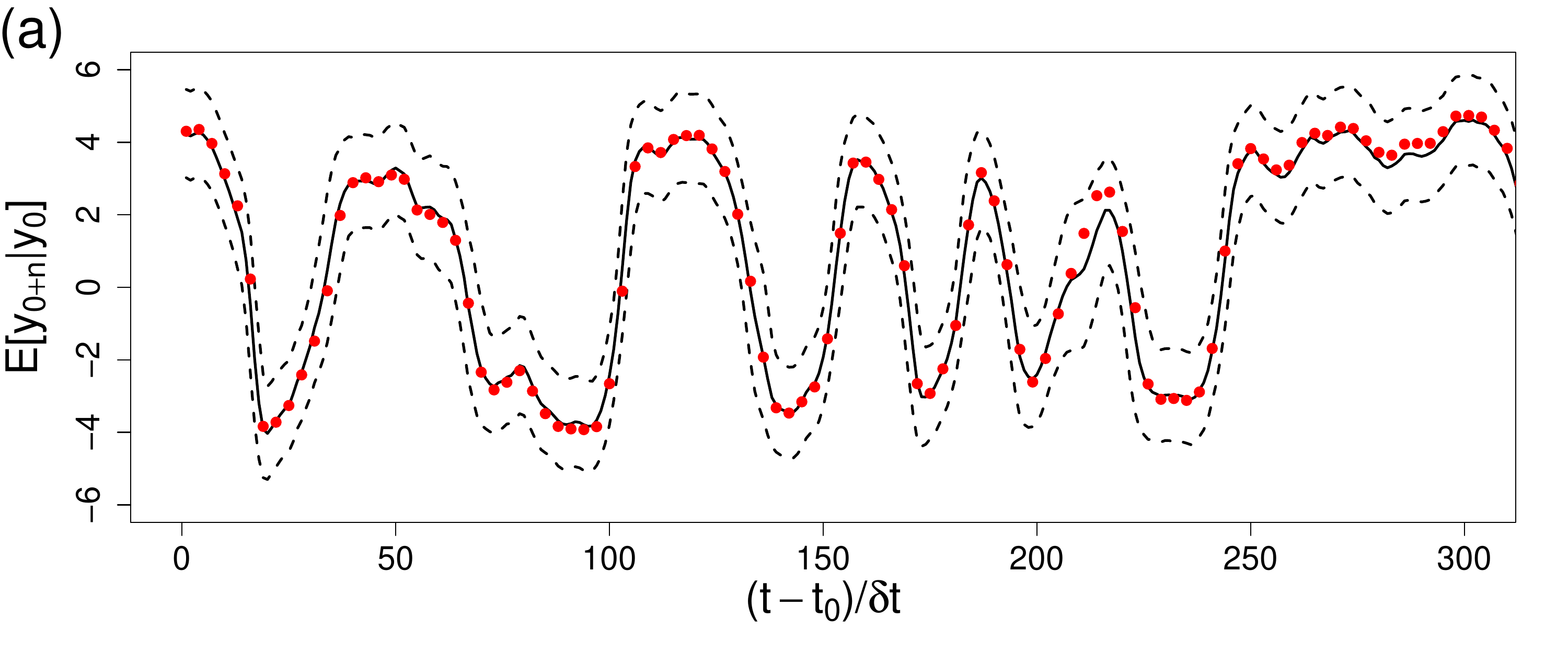}\\
  \includegraphics[width=0.95\textwidth]{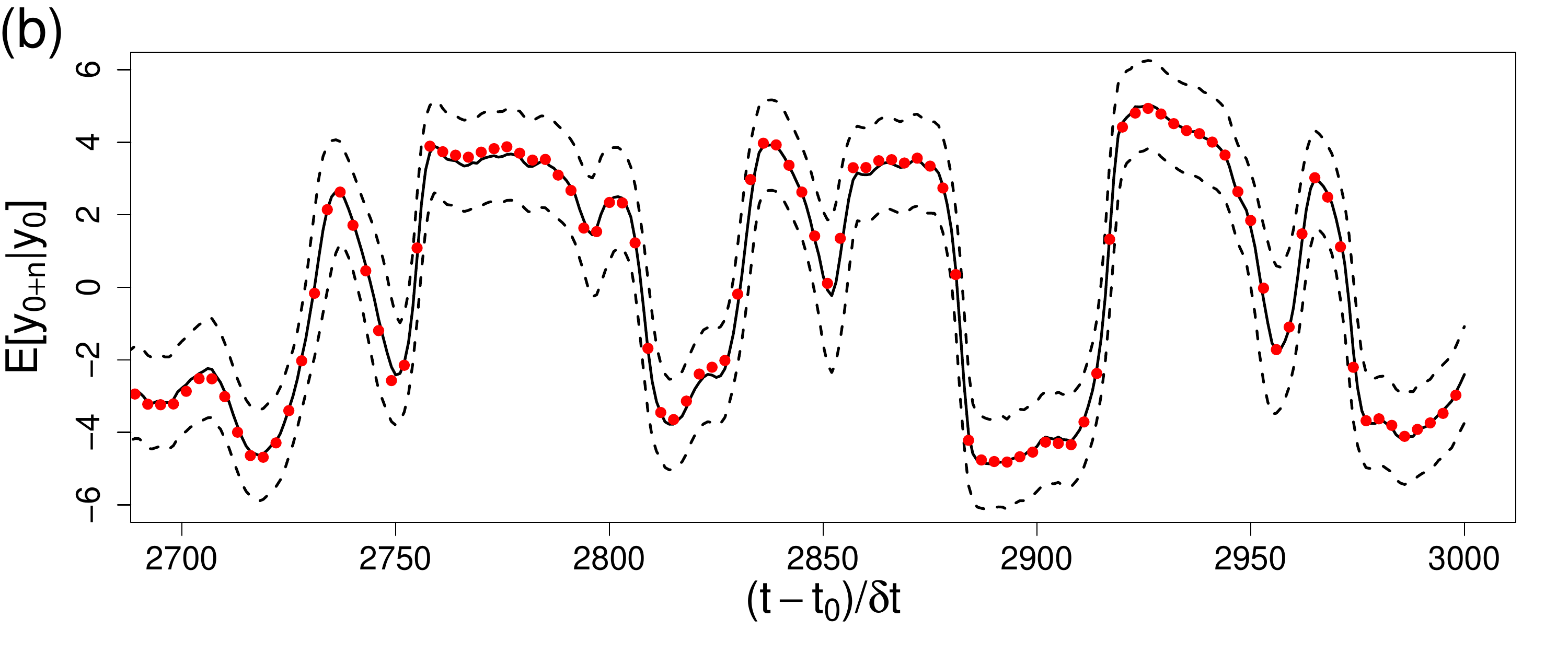}\\
  \includegraphics[width=0.95\textwidth]{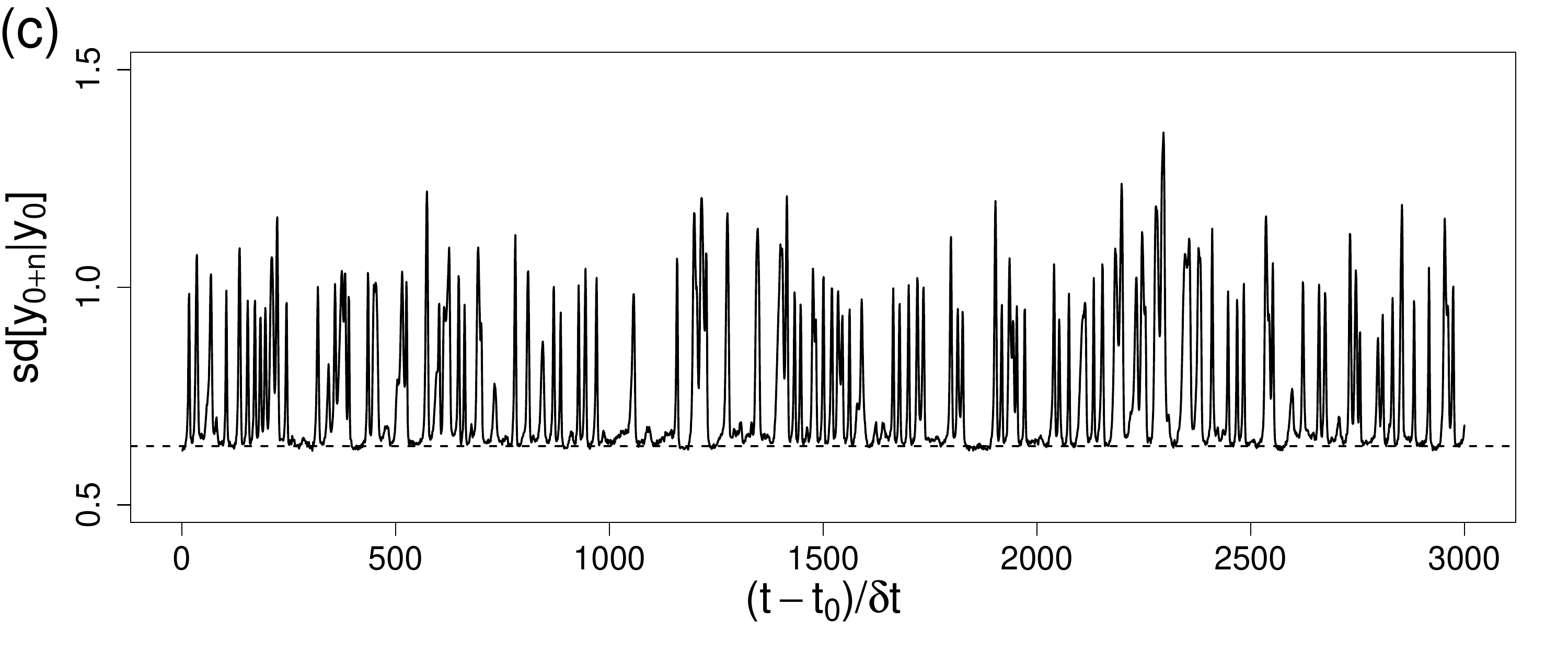}
  \caption{3000-step forecast by DE-LSTM for the exogenous forcing with $\theta = 1$. (a) and (b) show the DE-LSTM forecast at two different time windows. In (a,b), the solid and dashed lines denote the expectation, $E[\yhat_{t_0+n}|\yhat_{t_0}]$, and the 95\% confidence interval, respectively. The ground truth, $y(t)$, is shown as solid circles. (c) The predicted standard deviation $sd[\yhat_{t_0+n}|\yhat_{t_0}]$. The dashed line is the noise level, $\rho = 0.2sd[\yhat]$.}\label{fig:VDP_multi_step}
\end{figure}

Figure \ref{fig:VDP_multi_step} (a) shows a multiple-step forecast for the new VDP data, i.e., $\theta = 1$. The multiple-step prediction is performed for $t\in(0,600)$, which corresponds to 3,000 $\delta t$. It is surprising to observe that the 95\% confidence interval does not grow for this long-time forecast. Figure \ref{fig:VDP_multi_step} (b) shows the standard deviation of the predictive distribution as a function of forecast horizon. It is again shown that $sd[\yhat_{t_0+n}|\yhat_{t_0}]$ oscillates but does not grow in time. The coverage probability is 0.956, suggesting that prediction of the 95\% confidence interval is reliable. The Mackey-Glass dynamical system in section \ref{sec:MG} is chaotic, which makes it difficult to make a long time forecast. On the other hand, a Van der Pol oscillator has a unique limit cycle. The dynamics of the forced Van der Pol oscillator is determined by a competition between a restoring force to the limit cycle and the perturbation due to the exogenous forcing. Hence, a long time prediction is possible when the dynamics can be learned accurately from the data. 

\begin{figure}
  \centering
  \includegraphics[height=5cm]{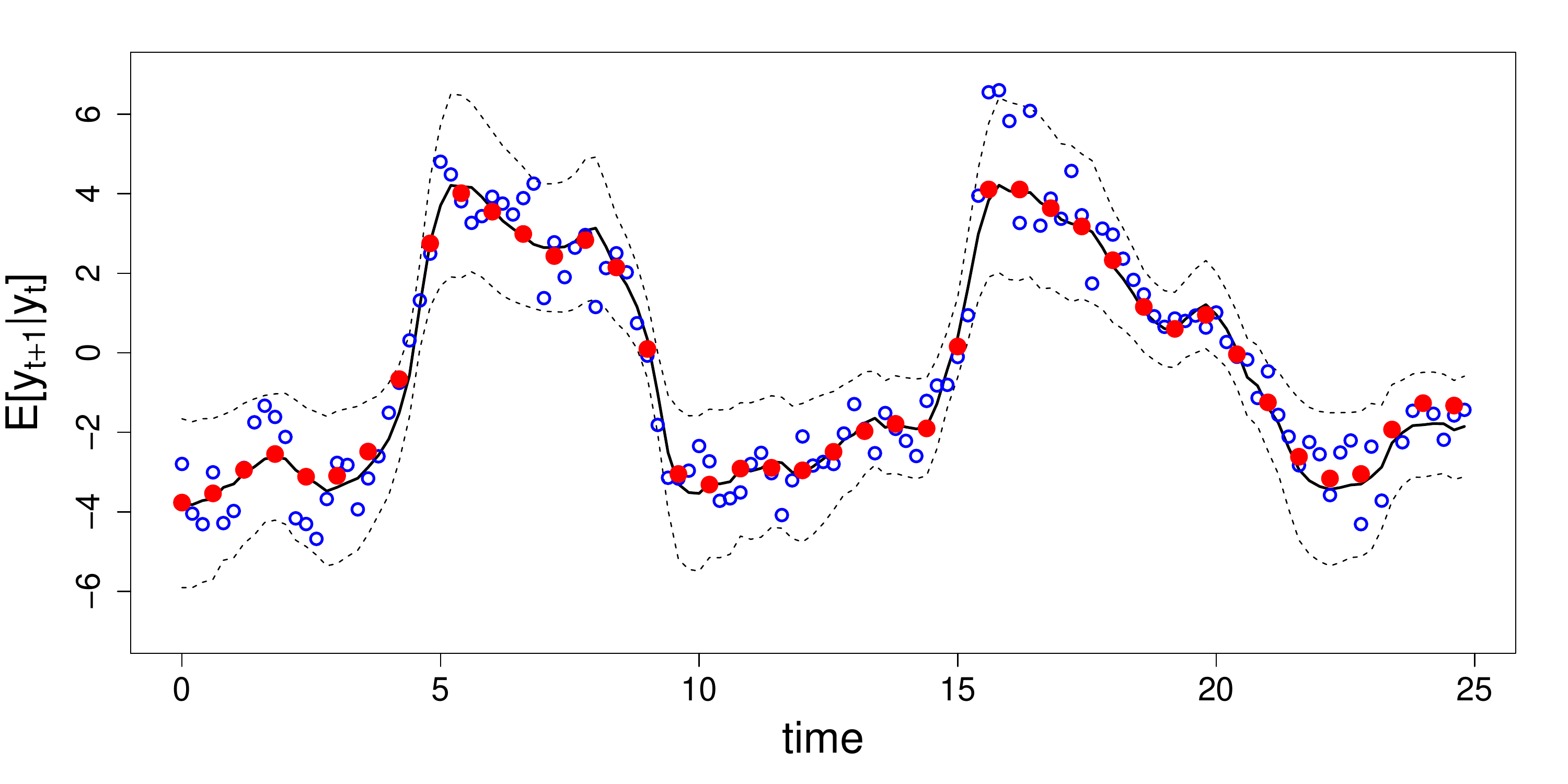}
  \caption{ Next-step prediction ($\frac{~~~}{~~~}$) and 95\% confidence interval ($\frac{~}{~}\frac{~}{~}\frac{~}{~}$) from DE-LSTM. The solid circles ({\color{red}$\bullet$}) denote the ground truth, $y(t)$, and the hollow circles ($\color{blue}\circ$) are the noisy observation, $\yhat_t$. }\label{fig:VDP_mul_noise_pred}
\end{figure}

To examine the robustness of DE-LSTM, a more complex noise is added to the ground truth as
\begin{equation} \label{eqn:VDP_mul_noise}
\yhat_t = y(t) + \epsilon_{1,t} + \epsilon_{2,t}.
\end{equation}
Here, $\epsilon_{1,t}$ and $\epsilon_{2,t}$ are independent Gaussian white noises ($\sim \mathcal{N}(0,\rho_t^2)$), of which standard deviations are $\rho_{1,t} = 0.1|y(t)|sd[y]$ and $\rho_{2,t} = 0.1sd[y]$. In other words, the noise is a combination of a multiplicative ($\epsilon_{1,t}$) and an additive ($\epsilon_{2,t}$) noise. 

Figure \ref{fig:VDP_mul_noise_pred} shows the next-step prediction by DE-LSTM for the complex noise (\ref{eqn:VDP_mul_noise}). Because of the multiplicative nature, the noise varies from 10\% to 70\% of $sd[y]$. It is shown that DE-LSTM still makes a good prediction for the multiplicative noise. The normalized root mean-square errors are listed in table \ref{tbl:VDP_mul_error}. Unless a special model specifically designed for the multiplicative noise structure is used, ARIMA and KF do not consider the complex nature of the noise in (\ref{eqn:VDP_mul_noise}). Hence, $e_{sd}$'s for ARIMA and KF are much larger then those of a simple additive noise in table \ref{tbl:VDP_error}. However, it is shown that the accuracy of DE-LSTM does not change noticeably.

\begin{figure}
  \centering
  \includegraphics[height=5.5cm]{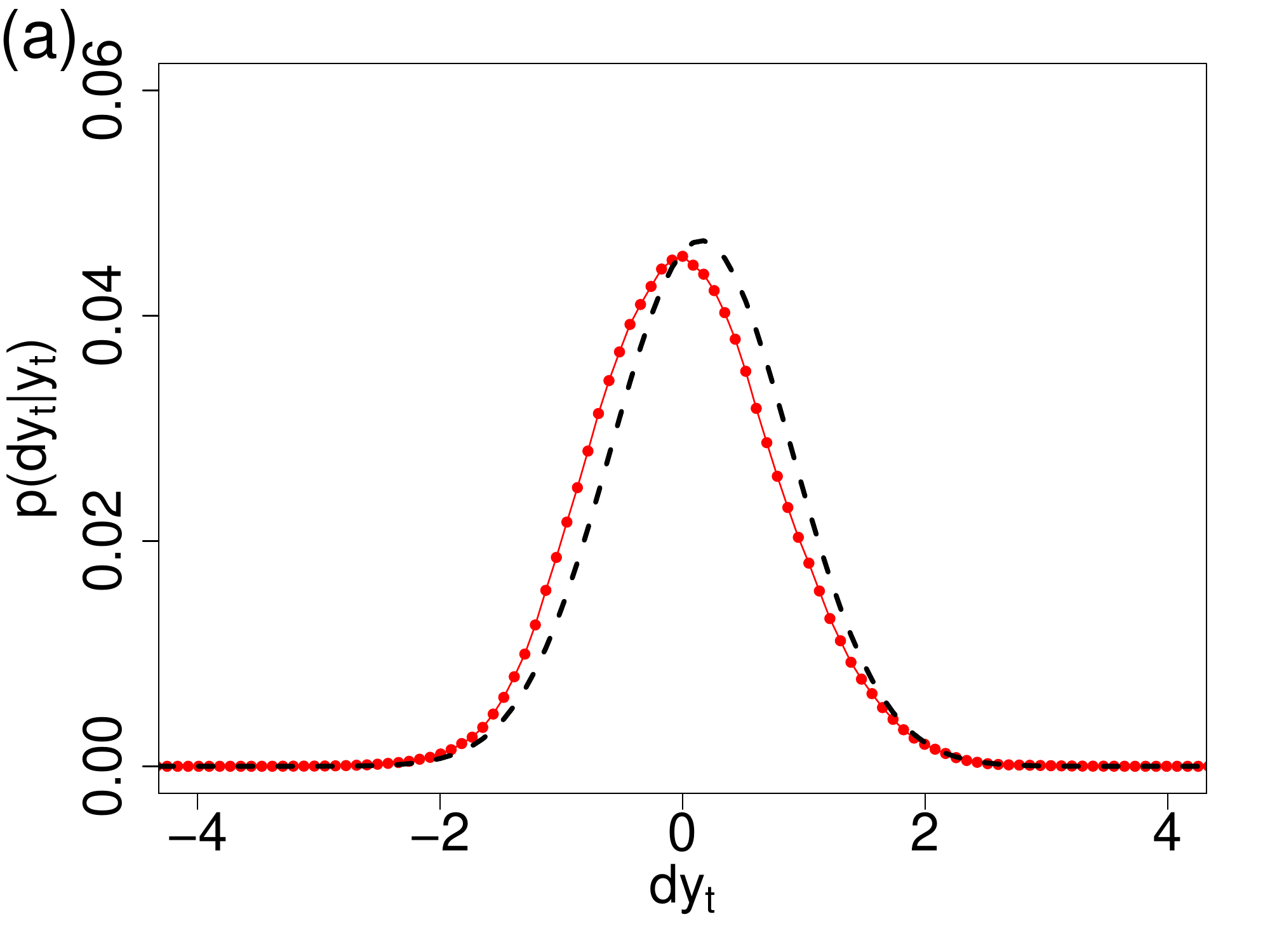}
  \includegraphics[height=5.5cm]{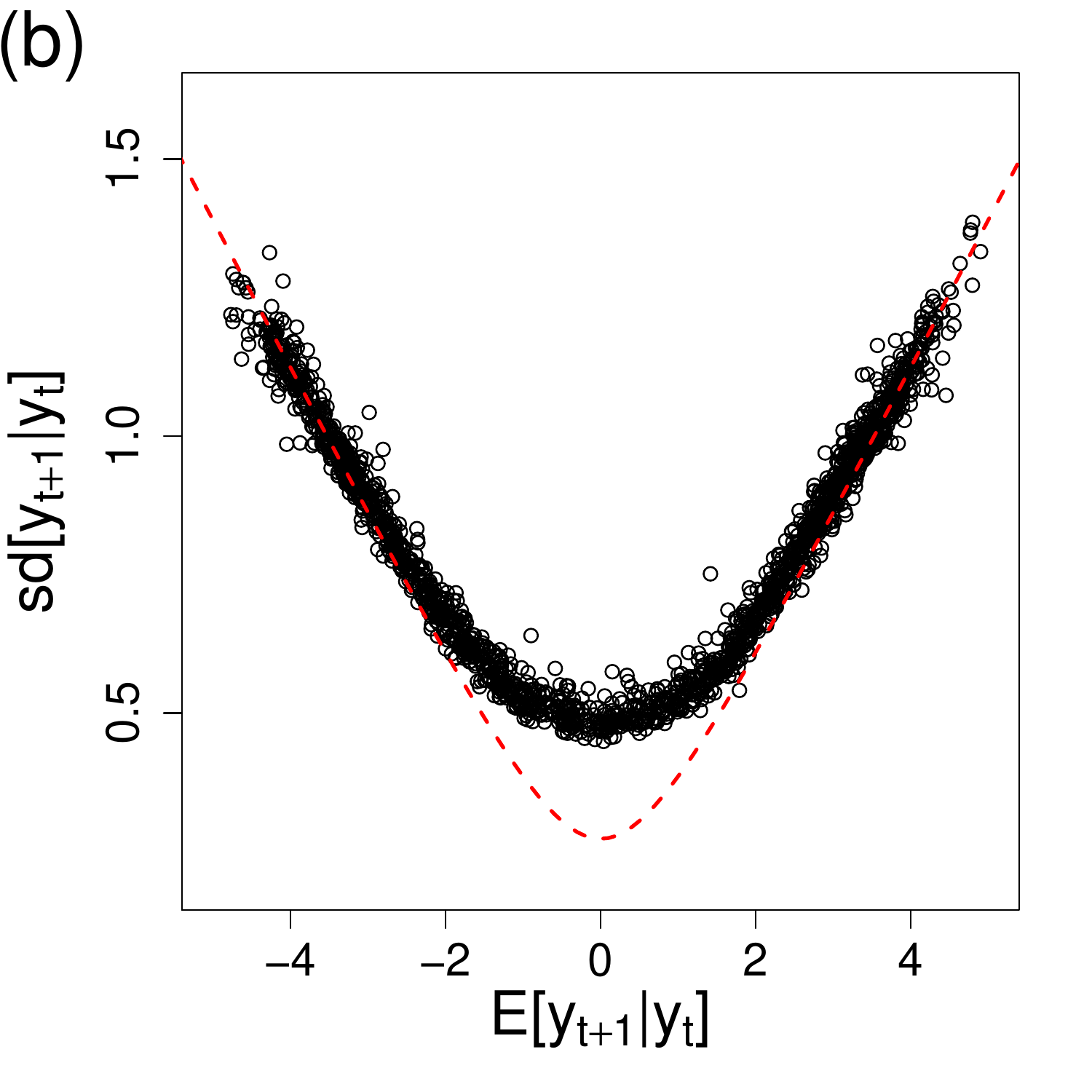}
  \caption{(a) The predictive probability distribution, $p(d\yhat_t|\yhat_t)$. The solid circles denote the DE-LSTM prediction and the dashed line is the ground truth. (b) The predicted standard deviation as a function of the expectation. The dahsed line in (b) is the ground truth, i.e., $sd[\epsilon_{1,t}+\epsilon_{2,t}|\yhat_t]$. } \label{fig:VDP_dist}
\end{figure}

\begin{table}
\center{
\caption{Normalized root mean-square errors for DE-LSTM, autoregressive integrated moving average (ARIMA), and Kalman filter (KF). DE-LSTM is trained with $\delta y = 0.03 sd[\yhat]$ and $\lambda = 10^{-2}$.} \label{tbl:VDP_mul_error}
\begin{tabular}{c|ccc}
\hline \hline 
&DE-LSTM & ARIMA & KF\\
\hline
$e_\mu$ & 0.26 & 1.17 & 1.14\\
$e_{sd}$ & 0.17 & 0.89 & 0.91\\
\hline\hline
\end{tabular}
}
\end{table}

Figure \ref{fig:VDP_dist} (a) shows the predictive probability distribution from DE-LSTM and the truth probability distribution of the noise. The snapshot is taken for a large noise case. It is shown that the predictive distribution approximates the ground truth very well. 

Figure \ref{fig:VDP_dist} (b) shows the predicted standard deviation as a function of the expectation. In the same plot, the ground truth is also shown; $sd[\epsilon_{1,t}+\epsilon_{2,t}|y_t] = 0.1sd[y]\sqrt{1+y_t^2}$. For a small value of $y(t)$, $-1 < E[\yhat_{t+1}|\yhat_t] < 1$, DE-LSTM overestimates the noise, while, for large noise levels, the standard deviation of DE-LSTM approximates the true noise very well. One of the possible reasons for the discrepancy at small $\yhat(t)$ is an imbalance in data. The data in the range of $|y(t)|<2$ is only 30\% of the total number of the data, while the data in $2<|y(t)|<4$ constitutes 61\% of the total.

\section{Concluding remarks} \label{sec:conclusions}
{
In this study, a deep learning algorithm is presented to predict the probability distribution of a noisy nonlinear dynamical system and to compute the time evolution of the probability distribution for a multiple-step forecast. The proposed deep learning algorithm employs the Long Short-Term Memory network to model the nonlinear dynamics and a softmax layer is used as a discrete approximation to the probability density function of the noisy dynamical system. The Long Short-Term Memory network provides a ``state space model'', of which internal state space consists of $N_c$-dimensional relaxation processes, in which $N_c$ is the number of the LSTM internal states. The relaxation timescales and forcing functions of the internal states are computed by the artificial neural network. A penalized log likelihood, or a regularized cross-entropy loss, is proposed to impose a smoothness condition in the predicted probability distribution. 
}

In most of the conventional time series models, the structure of the noise is assumed to be known, e.g., additive Gaussian noise, and the parameters of the given distribution are estimated from the data. However, in DE-LSTM, we only make an assumption of the smoothness of the probability density function and do not constrain the model to a specific probability distribution.  DE-LSTM is first tested against the Ornstein-Uhlenbeck process. It is shown that, without explicitly providing information on the underlying distribution, DE-LSTM can make a good prediction of the stochastic process. As expected, the regularized cross-entropy loss leads to a smooth probability distribution compared to DE-LSTM trained with the standard cross-entropy loss. In the numerical experiments against noisy nonlinear dynamical systems, it is shown that DE-LSTM can make a good prediction of the probability distribution when the noise has a more complicated structure, e.g., sum of multiplicative and additive noises, or Laplace noise.

{
We note that the internal states of LSTM are random variables when modeling a stochastic process. From the observation, we formulated the time evolution of the probability distribution of the noisy dynamical system as a high-dimensional integration of the transition probability of the internal states, $p(\bm{H}_{t+1}|\bm{H}_t)$, and proposed a Monte Carlo method for a multiple-step forecast. 
It is found that the prediction uncertainty of DE-LSTM dynamically adjusts over the forecast horizon. The prediction uncertainty may increase or decrease following the dynamics of the system. For the Mackey-Glass time series, the standard deviation of the multiple-step prediction grows at first, and then saturates for 1,000-step forecast. On the other hand, for a forced Van der Pol oscillator, it is shown that the standard deviation of the prediction does not grow even for 3,000-step forecast. In both cases, the coverage probability of 95\% confidence interval is about 0.94 $\sim$ 0.96, indicating DE-LSTM makes a reliable prediction of the uncertainty.
}

\section*{References}

\bibliographystyle{elsarticle-num}
\bibliography{LSTM_ref}

\end{document}